\documentclass[sn-mathphys]{sn-jnlAMS}% Math and Physical Sciences Reference Style
\usepackage{bm}
\usepackage{amsmath,subcaption}
\usepackage{paralist}
\jyear{2021}%

\theoremstyle{thmstyleone}%
\theoremstyle{thmstyletwo}%

\theoremstyle{thmstylethree}%

\usepackage{verbatim}

\renewcommand{\e}{\bm{e}}
\renewcommand{\f}{\bm{f}}
\renewcommand{\C}{\mathbf{C}}
\renewcommand{\x}{\bm{x}}
\renewcommand{\y}{\bm{y}}

\newcommand{\ave}[1]{{\left\langle #1 \right\rangle}}
\newcommand{\n}{\bm{n}}
\newcommand{\m}{\bm{n}_0}
\newcommand{\I}{\mathbf{I}}
\newcommand{\Lc}{\mathbf{L}}
\newcommand{\Lr}{\mathbf{L}_0}
\newcommand{\Qc}{\mathbf{Q}}
\newcommand{\Qr}{\mathbf{Q}_0}
\newcommand{\lrpar}{\ell_{0\parallel}}
\newcommand{\lrper}{\ell_{0\perp}}
\newcommand{\lcper}{\ell_{\perp}}
\newcommand{\lcpar}{\ell_{\parallel}}
\newcommand{\kv}{\bm{k}}
\newcommand{\ratetc}{r_\mathrm{t\to c}}
\newcommand{\ratect}{r_\mathrm{c\to t}}
\newcommand{\ts}{\emph{trans}}
\newcommand{\cs}{\emph{cis}}
\newcommand{\uv}{\bm{u}}
\newcommand{\vv}{\bm{v}}
\newcommand{\sphere}{\mathbb{S}^2}
\newcommand{\disk}{\mathbb{S}^1}
\newcommand{\Rr}{\bm{R}_0}
\newcommand{\Rc}{\bm{R}}
\newcommand{\Nt}{N_\mathrm{t}}
\newcommand{\Nc}{N_\mathrm{c}}
\newcommand{\curvature}{(\nablas\normal)}
\newcommand{\cframe}{(\e_1,\e_2,\e_3)}

\newcommand{\nablas}{\nabla\!_\mathrm{s}}

\newcommand{\tr}{\operatorname{tr}}
\newcommand{\daw}{\operatorname{daw}}

\newcommand{\F}{\mathbf{F}}
\newcommand{\trans}{^\mathsf{T}}
\newcommand{\Cf}{\C_{\f}}

\newcommand{\slab}{\mathsf{S}}
\newcommand{\surface}{\mathscr{S}}
\newcommand{\normal}{\bm{\nu}}
\newcommand{\dd}{\mathrm{d}}
\newcommand{\vt}{\vartheta}
\newcommand{\vp}{\varphi}
\newcommand{\nay}{(\nabla\y)}

\newcommand{\Ie}{$I_\mathrm{e}$}

\begin{document}

\title[Photoresponsive Nematic Elastomers]{Model for Photoresponsive Nematic Elastomers}


\author*[1]{\fnm{Andr\'e M.} \sur{Sonnet}}\email{andre.sonnet@strath.ac.uk}

\author*[2]{\fnm{Epifanio G.} \sur{Virga}}\email{eg.virga@unipv.it}
\equalcont{These authors contributed equally to this work.}

\affil*[1]{\orgdiv{Department of Mathematics and Statistics}, \orgname{University of Strathclyde}, \orgaddress{\street{26 Richmond Street}, \city{Glasgow}, \postcode{G1 1XH}, \country{U.K.}}}

\affil*[2]{\orgdiv{Dipartimento di Matematica}, \orgname{Universit\`a di Pavia}, \orgaddress{\street{Via Ferrata 5}, \city{Pavia}, \postcode{27100}, \country{Italy}}}


\abstract{We study the equilibria of a photoresponsive nematic elastomer ribbon within a continuum theory that builds upon the statistical mechanics model put forward by Corbett and Warner [Phys. Rev. E {\textbf{78}}, 061701 (2008)]. We prove that the spontaneous deformation induced by illumination is \emph{not} monotonically dependent on the intensity {\unboldmath $I$}. The ribbon's deflection first increases with increasing {\unboldmath $I$}, as expected, but then decreases and abruptly ceases altogether at a critical value {\unboldmath \Ie} of {\unboldmath $I$}. {\unboldmath \Ie}, which is enclosed within a hysteresis loop, marks a first-order \emph{shape transition}. Finally, we find that there is a critical value of the ribbon's length, depending only on the degree of cross-linking in the material, below which \emph{no} deflection can be induced in the ribbon, no matter how intense is the light shone on it.}




\maketitle

\section{Introduction}\label{sec:intro}
Nematic elastomers are elastic materials comprising cross-linked polymer networks made of nematogenic, rod-like molecules which at sufficiently low temperatures develop a collective orientational order. Despite polymer strands being cross-linked, the constituting monomers are quite loose; they are the fluid component of a mixture whose other component is a solid-like matrix kept together by the cross-linking bonds \cite{corbett:photomechanics}.

These materials can be \emph{reversibly} activated by changing the temperature across the nematic-isotropic transition. Upon increasing the temperature, the nematic order is decreased, the fluid becomes isotropic and the larger availability of orientational states produces a mechanical contraction of the solid matrix along the nematic director designating the pre-existing average molecular orientation. Conversely, decreasing the temperature across the nematic-isotropic transition, an elongation takes place along the newly reconstituted nematic director, as molecules would tend to be mostly oriented in that direction. This is perhaps the most remarkable mechanical property of nematic liquid crystal elastomers (LCEs): they can undergo a shape change of up to $400\%$ in a relatively narrow temperature range (including the nematic-isotropic transition of the fluid component).

Such a thermal activation mechanism was the only one known and studied until the groundbreaking work \cite{finkelmann:new} was published in 2001. That paper explored a new possible avenue for mechanical activation of nematic LCEs: using light instead of heat. The idea was simple: if a macroscopic change in shape is caused by a change in molecular order, the latter should result in the former, whatever means are employed. Now, order can either be decreased by raising the temperature or by disturbing molecules otherwise, making it harder for them to be oriented alike.

\begin{figure}[h]
	\centering
	\begin{subfigure}[t]{.5\linewidth}
		\centering
		\includegraphics[width=\linewidth]{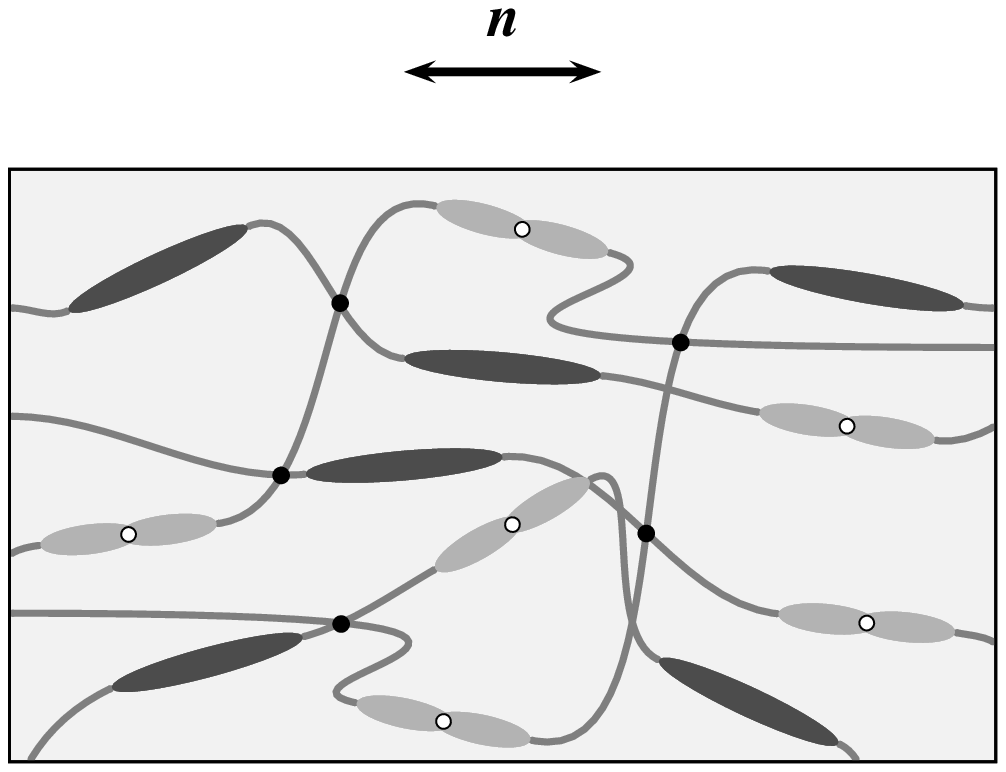}
		\caption{Photoresponsive molecules are in their \emph{trans}-state. The macroscopic director $\n$ is shown as a double-headed vector.}
		\label{fig:trans}
	\end{subfigure} 
	\quad
	\begin{subfigure}[t]{.4\linewidth}
		\centering
		\includegraphics[width=\linewidth]{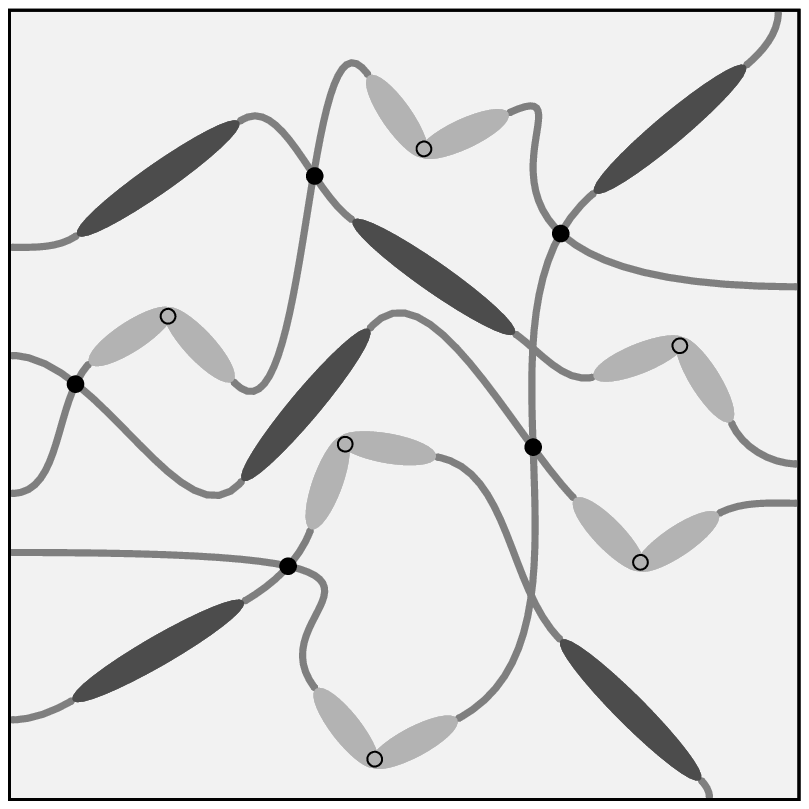}
		\caption{Photoresponsive molecules are in their \emph{cis}-state. The system is thought of as being isotropic: no macroscopic director can be defined.}
		\label{fig:cis}
	\end{subfigure}
	\caption{Cartoons illustrating the photoactivation of isomerizable molecules. They are straight in their \emph{trans}-state and bent in their \emph{cis}-state. Forward activation is induced by a photon absorption; backward relaxation is induced by thermal agitation, with no change in temperature involved. In our model, photoresponsive molecules are part of the polymer chains, within which, upon activation, they deplete nematic order. The case envisioned here is listed as case \eqref{item:case_two} in the text.}
	\label{fig:cartoon}
\end{figure}

Since the pioneering work of Eisenbach~\cite{eisenbach:isomerization}, this could be achieved by dispersing in the  material photoisomerizable molecules, such as azobenzene and other dyes, which undergo a \emph{trans-cis} isomerization upon absorbing a photon of appropriate frequency. These molecules, which are typically rod-like in the \emph{trans}-state, become bent in the \emph{cis}-state. Such a change in shape has a disrupting effect on surrounding molecules in the nematic phase, which remain straight, decreasing their degree of order (as first shown in \cite{stolbova:calculation}).\footnote{Similar effects could also be imparted on the director $\n$, but they will be ignored in this paper, as here $\n$ will be enslaved to the macroscopic deformation, an assumption which will be further discussed and justified below.} The specific situation envisioned in this paper is illustrated in Fig.~\ref{fig:cartoon}, which also shows dye molecules in both  \emph{trans}- and \emph{cis}-states.

Such a disruption of the nematic order is reversible. Photoresponsive molecules do not stay indefinitely in the \cs-state; this, although locally stable, has greater energy than the \emph{trans}-state, and thermal relaxation suffices to overcome the energy barrier that prevents exited molecules to drop to the \emph{trans}-state right away. Once photoresponsive molecules are back in the \emph{trans}-state, their reacquired rod-like shape no longer contrasts the alignment of monomers in the polymer strands, and nematic order can be reinstated. Thus, with no change in temperature, a change in order induced by light can produce a typical thermo-mechanical effect.

There are at least three possibilities for a photoisomerizable molecule to play its actuating  role within a nematic elastomer network: 
\begin{inparaenum}[(i)]
\item\label{item:case_one} by being freely dispersed through it, 
\item \label{item:case_two} by being part of the network itself, linked at a polymer chain at both its ends, 
\item \label{item:case_three} by being linked to a polymer chain side-wise at an end, with the other end dangling freely. 
\end{inparaenum}
They may all be effective.

Here, we shall consider case \eqref{item:case_two}: for simplicity, we shall further assume that when photoresponsive molecules are in the \ts-state they have the same length $a$ as the photoinert nematic monomers in the polymer strands. In their \ts-state, photoresponsive molecules are indistinguishable from nematogenic molecules, they obey the same statistics (see Fig.~\ref{fig:trans}). Light activation  induces the \emph{trans-cis} transition and changes the distance between the ends of the two arms of photoresponsive molecules (see Fig.~\ref{fig:cis}) from $a$ to $b<a$. This transition causes a depletion in the population of rods obeying nematic statistics and a repletion in the population of isotropically distributed rods: as such we regard photoresponsive molecules in the \cs-state. 

The stationary equilibrium between the light-induced \emph{trans-cis} transition and its reverse thermal relaxation determines the fraction $\phi$ of the \cs-population in terms of the nematic scalar order parameter $S$ and the orientation of the nematic director $\n$ relative to the wave polarization unit vector $\e$ \cite{corbett:nonlinear,corbett:polarization}. Both $\phi$ and $S$ in turn affect the \emph{step tensor} $\Lc$ describing the distribution of chain elements in a  representation of polymer strands as freely jointed rigid rods.

The intricate interplay between these microscopic processes is described by the model recalled in Sect.~\ref{sec:model}. This model is originally due to Corbett and Warner \cite{corbett:nonlinear,corbett:linear,corbett:polarization} (see also \cite{corbett:deep,corbett:photomechanics}).

In Sect.~\ref{sec:energy}, we build the macroscopic theory that we shall adopt here. Its main ingredient is the celebrated \emph{trace formula} for the elastic free-energy density (per unit volume) of nematic LCEs that has long been studied \cite{blandon:deformation,warner:theory,warner:elasticity,warner:nematic} (see also Chap.~6 of \cite{warner:liquid}).

As effectively recalled in \cite{white:photomechanical_collection}, nematic LCEs have also come to be known in the specialized literature under a variety of names, including liquid crystal polymers, cross-linked liquid crystal polymers, and liquid crystal polymer networks. What marks the difference between these names is their different range of applicability, which is essentially decided by the extent of cross-linking: the higher this is, the stiffer the material becomes and the more is the nematic director $\n$ linked to the polymer matrix. When $\n$ is completely enslaved to the macroscopic deformation, which is the case of extreme cross-linking, also the name \emph{nematic polymer network} (NPN) is used for these materials.\footnote{As well as nematic \emph{glass}, as for example in \cite{modes:disclination}.}

It has recently become clear \cite{ware:programmed} that the mechanical response changes continuously with the extent of cross-linking. To fix ideas, we find at the NPN end of the spectrum a transition temperature ranging from $60$ to $100\, ^\circ\mathrm{C}$ with shear modulus parallel to the nematic director in the range of $1$--$2\,\mathrm{GPa}$. At the opposite end of the spectrum transition temperatures are below $25\, ^\circ\mathrm{C}$ with shear moduli about $100\,\mathrm{MPa}$ or lower \cite{white:photomechanical_collection}.

Since in our model the scalar order parameter $S$ is \emph{not} the driving parameter of spontaneous deformation, as the temperature is kept fixed, a further potential must supplement the trace formula, which penalizes departures from the equilibrium value $S_0$ of $S$, which is dictated by temperature. This role will be played here (as was in \cite{bai:photomechanical}) by the Maier-Saupe potential \cite{maier:einfache}.

In Sect.~\ref{sec:energy}, we shall also adapt to the present setting the surface elastic free energy density (per unit area) obtained in \cite{ozenda:blend} for a thin NPN sheet by extending a method of dimension reduction, which is standard in the theory of plates and also known as the Kirchhoff-Love hypothesis \cite{ozenda:kirchhoff}. This surface free energy features both stretching and bending contributions, which here are reformulated in the language of the model illustrated in Sect.~\ref{sec:model}.

In Sect.~\ref{sec:ribbon}, we apply the reduced elastic free energy for a sheet to a narrow ribbon and study its equilibrium configurations in terms of a dimensionless intensity parameter $I$. Resorting to a uniformity approximation, we simplify the total free-energy functional for a ribbon to an extent that makes it possible to find its critical points in closed form. The bifurcation analysis that ensues reveals a non-monotonic dependence on $I$ of the maximum deflection angle of an illuminated ribbon.

Our conclusions are collected and discussed in Sect.~\ref{sec:conclusion} together with comparisons of our work with that of others. The paper is closed by an Appendix, where for completeness we recall the reasoning that is followed in \cite{corbett:linear} to justify both the equilibrium value $\phi$ of the \cs-population and the expressions for the principal chain steps (i.e., the eigenvalues of $\Lc$) after photoactivation.

A vast, nearly intimidating literature is available on nematic LCEs. The classical reference is the influential book by Warner and Terentjev~\cite{warner:liquid}; the theoretical literature that preceded and prepared for it \cite{blandon:deformation,warner:soft,terentjev:orientation,verwey:soft,verwey:multistage,verwey:elastic} is also of interest. General continuum theories have also been proposed \cite{anderson:continuum,zhang:continuum,mihai:nematic}, some also very recently. Applications are abundant; a collection can be found in a book \cite{white:photomechanical_book} and a recent special issue \cite{korley:introduction}. Finally, the interested reader can gain some valuable guidance from the reviews \cite{mahimwalla:azobenzene,ube:photomobile,white:photomechanical,ula:liquid,pang:photodeformable,kuenstler:light,warner:topographic}.

The specific theme of this paper is photoactivation of NPN sheets; the recent studies \cite{bai:photomechanical,korner:nonlinear,goriely:rod} are closely related to it and, notwithstanding the differences, they have been inspirational to us.
 
\begin{comment}
\subsection{Phenomenological account on these material, highlighting the difference between thermal and optical effects.}
\subsection{Summary of the vast literature on the subject.}
\subsection{Plan of the paper.}
\end{comment}
\section{The Corbett-Warner Model}\label{sec:model}
\begin{comment}
\subsection{Recap of the model, highlighting the main assumptions on which it rests.}

\subsection{It might be instructive to sketch the derivation of the equilibrium condition that determines the cis fraction phi.}

\subsection{Distinction between polarized and unpolarized light.}
\end{comment}
In this section, mainly following \cite{corbett:nonlinear,corbett:linear,corbett:polarization}, we present a statistical mechanics model put forward by Corbett and Warner to describe the interaction of an incoming polarized light wave with a nematic elastomer containing photoactivable molecules in its polymer strands, as depicted in Fig.~\ref{fig:cartoon}. For completeness, the reasoning that led these authors to their understanding of shape effects induced on polymer networks by the \ts-\cs\ transition is described in more detail in Appendix~\ref{sec:appendix}. Here, our main focus is on the foundations of the model and its main outcomes, which will form the basis for our macroscopic theory laid down in the following section.

The model is based on the following assumptions:
\begin{enumerate}
	\item Photoresponsive molecules in their \ts-state and photoinert (non-photoresponsive) molecules are assumed to be \emph{statistically} identical.
	\item Photoresponsive molecules in the \cs-state are statistically \emph{isotropic}.
	\item The forward \ts-\cs\ photoisomerization is powered by light, whereas the backward \cs-\ts\ transition is spontaneously driven by thermal agitation.
	\item The \ts-\cs\ reaction is treated at the \emph{single-molecule} level: equilibrium at one molecule is not affected by its interaction with surrounding molecules. 
\end{enumerate}
 
Polymer strands, constituted of both photoresponsive and photoinert molecules, are represented as chains of freely jointed rigid rods; the orientation in space of an individual rod will be represented by a unit vector $\uv\in\sphere$. The order tensor $\Qc$ describing the alignment of mesogenic monomers is defined as
\begin{equation}
	\label{eq:Q_definition}
	\Qc:=\ave{\uv\otimes\uv-\frac13\I},
\end{equation}
where $\I$ is the identity tensor and the brackets $\ave{\cdots}$ designate ensemble averaging. We shall assume that $\Qc$ is \emph{uniaxial}, and so it can be
represented as
\begin{equation}
	\label{eq:Q_unixial_representation}
	\Qc=S\left(\n\otimes\n-\frac13\I\right),
\end{equation} 
where $S$ is the \emph{scalar} order parameter, ranging in the interval $[-\frac12,1]$, and $\n\in\sphere$ is the nematic \emph{director}.\footnote{$\Qc$ is the deviatoric part of the second-moment distribution of monomer $\uv$'s: $S=-\frac12$ when all $\uv$'s are uniformly distributed in the plane orthogonal to $\n$, while $S=1$ when all $\uv$'s are aligned with $\n$, and $S=0$ when all $\uv$'s are isotropically distributed (and $\n$ is undefined), see, for example, Chapt.~1 of \cite{virga:variational}.} It readily follows from \eqref{eq:Q_unixial_representation} that 
\begin{equation}
	\label{eq:S_P_2}
	S=\ave{P_2(\uv\cdot\n)},
\end{equation}
where $P_2$ is the second Legendre polynomial.

In the following, $S$ and $\n$ will represent the scalar order parameter and the nematic director in the \emph{present} (photoactivated) configuration of the material, after the spontaneous deformation of the body induced by illumination has taken place. Since we also assume that prior to photoactivation the polymer system had been cross-linked in the nematic phase, we shall denote by $S_0$ and $\m$ the scalar order parameter and the nematic director in the \emph{reference} configuration, prior to illumination. $\Qr$, related to $S_0$ and $\m$ as $\Qc$ is to $S$ and $\n$ in \eqref{eq:Q_unixial_representation}, will denote the corresponding order tensor. 

The nematic order in the material is also reflected by the \emph{step tensor}, which describes the spatial organization of polymer strands (see Appendix~\ref{sec:step_tensors} for a formal definition). Since there are two different polymer  organizations, in the reference and present configurations, there will correspondingly be two step tensors, $\Lr$ and $\Lc$. They have the same uniaxial form as $\Qr$ and $\Qc$, respectively, and are represented as
\begin{subequations}	\label{eq:step_tensors}
\begin{eqnarray}
\Lr&=&\lrper\I+(\lrpar-\lrper)\m\otimes\m,\label{eq:step_tensor_reference_text}\\
 \Lc&=&\lcper\I+(\lcpar-\lcper)\n\otimes\n,\label{eq:step_tensor_current_text}
\end{eqnarray} 
\end{subequations}
where $(\lrper,\lrpar)$ and $(\lcper,\lcpar)$ are the corresponding \emph{principal chain steps} in the directions orthogonal and parallel to the directors $\m$ and $\n$ (see Appendix~\ref{sec:step_tensors}).

The principal chain steps $(\lrper,\lrpar)$ in the reference configuration are related to the scalar order parameter $S_0$ in the following way,
\begin{equation}
	\label{eq:principal_chain_steps_reference}
	\lrper=a(1-S_0),\quad \lrpar=a(1+2S_0),
\end{equation}
where $a$ is the \emph{step length} that both photoinert nematogenic molecules  and photoresponsive ones in the \emph{trans} configuration are  assumed to have in common. 

A statistical mechanics argument of Corbett and Warner \cite{corbett:nonlinear,corbett:polarization} justifies writing the principal chain steps $(\lcper,\lcpar)$ in the present configuration as
\begin{equation}\label{eq:principal_chain_steps}
	\lcper=a\left[(1-\phi)(1-S)+\phi\left(\frac{b}{a}\right)^2 \right],\quad \lcpar=a\left[(1-\phi)(1+2S)+\phi\left(\frac{b}{a}\right)^2 \right],
\end{equation}
where $S$ is the scalar order parameter after photoactivation, $b<a$ is the step  length of photoresponsive nenatogens in the \emph{cis} configuration, and $\phi$ is the \emph{number fraction}  of these molecules (see Appendix~\ref{sec:step_tensors}). It should be noted that for $\phi=0$ equation \eqref{eq:principal_chain_steps} reduces to \eqref{eq:principal_chain_steps_reference}, only with $S$ instead of $S_0$.

The equilibrium value of $\phi$ depends on how light impinges on the material. Letting the unit vector $\e\in\sphere$ denote the polarization of the  incoming light, that is, the direction of vibration of the electric field, it was shown in \cite{corbett:polarization} that in equilibrium
\begin{equation}\label{eq:phi_equation}
	\phi=A\frac{\mathcal{I}[1+S(3(\n\cdot\e)^2-1)]}{3\mathcal{I}_\mathrm{c}+\mathcal{I}[1+S(3(\n\cdot\e)^2-1)]},
\end{equation}	
where $A$ is the fraction of photoresponsive nematogens in a polymer strand, $\mathcal{I}$ is the intensity of light, and $\mathcal{I}_\mathrm{c}$ a characteristic intensity, related to both forward and reverse isomerization rates (see Appendix~\ref{sec:cis_population}). 

For unpolarized light, which is the case that we shall consider in this paper, $(\n\cdot\e)^2$ should be replaced in \eqref{eq:phi_equation} by its average over a uniform distribution of $\e$ in the unit circle $\disk_{\kv}$ lying in the plane orthogonal to the unit vector $\kv\in\sphere$ designating the direction of propagation of light. Since
\begin{equation}
	\label{eq:average_S_1}
	\ave{(\n\cdot\e)^2}_{\disk_{\kv}}=\n\cdot\ave{\e\otimes\e}_{\disk_{\kv}}\n=\n\cdot\frac12(\I-\kv\otimes\kv)\n=\frac12(1-(\n\cdot\kv)^2),
\end{equation}
equation \eqref{eq:phi_equation} will be replaced here by
\begin{equation}
	\label{eq:phi_equation_unpolarized}
	\phi=A\frac{I\left[1+\frac{S}{2}(1-3(\n\cdot\kv)^2)\right]}{3+I\left[1+\frac{S}{2}(1-3(\n\cdot\kv)^2)\right]},
\end{equation} 	
where
\begin{equation}
	\label{eq:relative_intensity}
	I:=\frac{\mathcal{I}}{\mathcal{I}_\mathrm{c}}
\end{equation}
is the \emph{relative} intensity, a dimensionless quantity that in our analysis will play the role of a \emph{control} parameter.

It is a simple matter to check that for $S<1$ the function delivering $\phi$ in \eqref{eq:phi_equation_unpolarized} tends to the asymptotic value $A$ for $I\to\infty$ (the bleaching limit). Moreover, letting $\vt$ be the angle that $\kv$ makes with $\n$, for a given intensity $I$, $\phi$ is either a monotonic increasing or decreasing function of $\vt$ for either $S>0$ or $S<0$, respectively. Thus, for positive $S$, the photoactivation mechanism described above is most efficient when $\n$ and $\kv$ are at right angles. If a deformation of the body moves $\n$ closer to $\kv$, at a given intensity, photoactivation may be weakened.

In general, light is absorbed in a material, in a way that depends on the penetration depth and direction of propagation of the radiation, as described, for example, by Beer's law \cite{beer:bestimmung} (see also Chapt.~1 of \cite{fox:otptical}). Here, this classical picture is further complicated by the fact that absorption also depends on the population of photoresponsive molecules in the \ts-state, but not on those in the \cs-state. To account properly for this would require coupling the population evolution equation resulting in \eqref{eq:phi_equation} at equilibrium with an attenuation equation for the intensity, as proposed in \cite{corbett:deep}.

Here, we shall only be concerned with thin films, and we shall make the approximation that the intensity of light remains unaffected through the thickness of the body. We shall refer to this as the \emph{photo-uniformity} approximation. Of course, this will have a price: we shall not be able to capture the symmetry breaking associated with the direction of propagation of light. Whatever instability we shall be able to predict with our continuum theory, flipping \emph{towards} or \emph{away} from light, such as in the experiments described in \cite{yu:directed}, \cite{liu:light}, or \cite{camacho-lopez:fast}, will come from extrinsic \emph{ad hoc} considerations. We shall be contented with capturing exactly (possibly in a closed form) just the \emph{flipping} (in either direction), if any.

\begin{comment}
\subsection{Step tensors in both reference (inactive) and present (activated) configurations.}
I write the (uniaxial) step tensors in both the reference and current configurations (before and after illumination, respectively) as
\subsection{Principal chain steps in both configurations. Should we say a few words about their derivation?}
We assume that the material is a \emph{nematic polymer network} (also known as a nematic glass), crosslinked in the nematic phase with scalar order parameter $S_0$ in the reference configuration, so that 
\end{comment}

\section{Free-Energy Functional}\label{sec:energy}
\begin{comment}
\subsection{Trace formula.}
\end{comment}
Our continuum theory is based on the \emph{trace formula} for the elastic free energy per unit volume (in the reference configuration) of nematic LCEs in the form put forward in \cite{finkelmann:elastic},
\begin{equation}
	\label{eq:free_energy_density}
	f_\mathrm{e}=\frac12\mu_0\bigg\{\tr(\F\Lr\F\trans\Lc^{-1})+\ln\left(\frac{\det\Lc}{\det\Lr}\right)\bigg\},
\end{equation}
where $\F$ is the deformation gradient and the \emph{shear} elastic modulus $\mu_0$ is given by
\begin{equation}
	\label{eq:mu_0}
	\mu_0=n_\mathrm{s}k_BT,
\end{equation}
in terms of the number density of polymer strands $n_\mathrm{s}$, the Boltzmann constant $k_B$, and the absolute temperature $T$.

Equation \eqref{eq:free_energy_density} is the nematic generalization of the classical elastic energy density for elastomers, which Rivlin \cite{rivlin:large_I} called \emph{neo-Hookian} (see also  \cite{kubo:large} and the detailed account in Sect.~95 of \cite{truesdell:non-linear_third}, which sets this constitutive law within the larger class of \emph{Mooney-Rivlin} materials),
\begin{equation}
	\label{eq:neo_Hookian_energy}
	f_0:=\frac12\mu_0\tr\Cf,
\end{equation}
where $\Cf:=\F\trans\F$ is the (three-dimensional) right Cauchy-Green tensor associated with a deformation $\f$. A statistical mechanics justification has also been derived for \eqref{eq:neo_Hookian_energy} from various theories of long chain molecules.\footnote{An exposition of statistical theories for rubber can be found in the landmark book \cite{treloar:non-linear_third}.}

The function in \eqref{eq:free_energy_density} is obtained by adapting the simplest realization of these theories \cite{deam:theory} to the case where in both the reference and present configurations the distribution of monomers in a polymer chain is anisotropic. 

When the principal chain steps $(\lrper,\lrpar)$ and  $(\lcper,\lcpar)$ in both the reference and present configurations are prescribed, as is the case where the corresponding scalar order parameters $S_0$ and $S$ are given as functions of temperature,\footnote{No photoactivation takes place in this case and a spontaneous deformation of the body is induced by a change in temperature, as recently considered, for example, in \cite{ozenda:blend,singh:model} and in many other studies reviewed in \cite{warner:topographic}.} the second term in \eqref{eq:free_energy_density} is not affected by the deformation and can be omitted, thus reducing \eqref{eq:free_energy_density} to the \emph{bare} trace formula of \cite{blandon:deformation} (also discussed in \cite{warner:new}), which depends only on $\F$ and $\n$.\footnote{An alternative theory building on the bare trace formula is presented in \cite{mihai:nematic}, where the existence of an isotropic reference configuration plays a central role.}

Our theory will be based on two further assumptions:
\begin{enumerate}[(a)]
	\item\label{item:assumption_1}
	We assume that the director $\n$ is \emph{enslaved} to the deformation, so that
\begin{equation}\label{eq:n_conveyed}
	\n=\frac{\F\m}{\vert\F\m\vert},
\end{equation}
which says that the nematic director $\m$ imprinted in the polymer network at the cross-linking time is \emph{conveyed} by the solid matrix of the body.
\item\label{item:assumption_2}
We assume that the material is \emph{incompressible}, so that $\F$ is subject to the constraint
\begin{equation}
	\label{eq:incompressibility}
	\det\F=1.
\end{equation}
\end{enumerate}

Clearly, assumption \eqref{item:assumption_1} is expected to be more realistic for \emph{strongly} cross-linked polymers than for \emph{weakly} cross-linked ones.\footnote{We shall see shortly below how this difference is accounted for by the choice of a dimensionless model parameter.} In real life, NPNs (or glassy nematics) are good examples of the former category, as are ordinary nematic LCEs of the latter. 
On the other hand, while ordinary nematic LCEs have Poisson ratio $\nu$ close to $1/2$, and so they are nearly incompressible, NPNs may have $\nu\in(1/2,2)$ \cite{van_oosten:glassy,warner:topographic}.

The above assumptions will simplify our analysis a great deal, making tractable the nonlinear problem discussed in the next section. We believe that our model is most appropriate for NPNs and that the conclusions reached here for these materials remain qualitatively valid, should either \eqref{eq:n_conveyed} or \eqref{eq:incompressibility} be partly relaxed, for example, by means of a penalizing potential.\footnote{However, relaxing \eqref{eq:n_conveyed} completely might ignite pattern formation at a fine scale (such as those studied in \cite{bai:photomechanical}), which might change the overall picture.}

It is a matter of laborious, but simple algebra, also making use of \eqref{eq:n_conveyed} and \eqref{eq:step_tensors}, to give $f_\mathrm{e}$ in \eqref{eq:free_energy_density} the following form
\begin{equation}
	\label{eq:free_energy_density_conveyed}
	\begin{split}
	f_\mathrm{e}=\frac12\mu_0\bigg\{&\frac{\lrper}{\lcper}\tr\Cf+\frac{1}{\lcpar}(\lrpar-\lrper)\m\cdot\Cf\m\\&+\lrper\left(\frac{1}{\lcpar}-\frac{1}{\lcper}\right)\frac{\m\cdot\Cf^2\m}{\m\cdot\Cf\m}+\ln\frac{\lcpar}{\lrpar}+2\ln\frac{\lcper}{\lrper}\bigg\}.
	\end{split}
\end{equation}

In our model, the scalar order parameter $S$ is not prescribed, but is free to adjust itself in response to illumination. The elastic free energy must then be supplemented by the bulk condensation energy for the nematic component. Here we depart from \cite{finkelmann:elastic}. While they adopted the Landau-deGennes phenomenological approach, we derive  the appropriate condensation potential  $f_\mathrm{c}$ (per unit volume) from the Maier-Saupe mean-field formulation of nematic condensation (from the isotropic phase), as described in Sect.~1.3 of \cite{sonnet:dissipative}. We set
\begin{equation}
	\label{eq:condensantion_energy}
	f_\mathrm{c}=n_\mathrm{n}k_BT(1-\phi)\psi_\mathrm{MS}(S),
\end{equation}  
where $n_\mathrm{n}$ is the number density of nematogenic molecules, here appropriately \emph{reduced} to account for their fraction in the \emph{cis}-state (which are no longer in the nematic phase), and 
\begin{equation}\label{eq:Maier-Saupe_potential}
	\psi_\mathrm{MS}(S) := J\left(\frac13 S^2 - \frac23 S\right) - \ln\left(\frac{\daw(\sqrt{JS})}{\sqrt{JS}}\right).
\end{equation}
In \eqref{eq:Maier-Saupe_potential}, $J$ is the 
Maier-Saupe molecular \emph{interaction energy} (scaled to $k_BT$) and $\daw$ denotes the Dawson integral, defined as
\begin{equation}
	\label{eq:Dawson_integral}
	\daw(x):=\mathrm{e}^{-x^2}\int_0^x\mathrm{e}^{t^2}\dd t\quad\text{for}\quad x\in\mathbb{R}.
\end{equation}
The minimizer of $\psi_\mathrm{MS}$ depends only on $J$, which will be chosen so that $\psi_\mathrm{MS}$ attains its minimum at $S=S_0$, the scalar order parameter prior to illumination.

Scaling the total energy density $f_\mathrm{t}:=f_\mathrm{e}+f_\mathrm{c}$ to $n_\mathrm{n}k_BT$ (so as to make it dimensionless),  we arrive at
\begin{equation}
	\label{eq:total_energy_density}
		\begin{split}
		f_\mathrm{t}=\frac12 \bigg\{\mu\bigg[&\frac{\lrper}{\lcper}\tr\Cf+\frac{1}{\lcpar}(\lrpar-\lrper)\m\cdot\Cf\m\\&+\lrper\left(\frac{1}{\lcpar}-\frac{1}{\lcper}\right)\frac{\m\cdot\Cf^2\m}{\m\cdot\Cf\m}+\ln\frac{\lcpar}{\lrpar}+2\ln\frac{\lcper}{\lrper}\bigg]\\&+2(1-\phi)\psi_\mathrm{MS}(S)\bigg\},
	\end{split}
\end{equation}
where 
\begin{equation}
	\label{eq:reduced_modulus}
	\mu:=\frac{n_\mathrm{s}}{n_\mathrm{n}}
\end{equation}
is the \emph{reduced} shear modulus. It is precisely the value of $\mu$ that describes in our model the degree of cross-linking in the material: the larger $\mu$, the stronger the cross-linking. Here, following \cite{corbett:polarization}, we shall identify two values of $\mu$ as representatives for two alternative regimes: $\mu=1/10$, for \emph{strongly} cross-linked polymers, and $\mu=1/50$, for \emph{weakly} cross-linked ones.

\begin{comment}
\subsection{High cross-linking density assumption, giving rise to nematic polymer networks (or nematic glasses). This will justify taking large values of mu (strong cross-linking).}

\subsection{Maier-Saupe potential according to the SV parameterization.}
\end{comment}

\subsection{Dimension reduction}
To treat (in the following section) the equilibrium of a thin sheet, we must first perform an appropriate reduction of the total free-energy density $f_\mathrm{t}$ (per unit volume) in \eqref{eq:total_energy_density} to a surface free-energy (per unit area) to be attributed to a surface in three-dimensional space representing the deformed shape of the sheet.

We accomplish this task following mainly \cite{ozenda:blend}.\footnote{See also \cite{singh:model} for the specific case of a thermally activated ribbon, which differs from the optically activated one studied in the following section.} We perform an expansion of $f_\mathrm{t}$ retaining up to the cubic terms in the sheet's thickness, thus identifying \emph{stretching} and \emph{bending} contents in the surface energy by the power in the thickness they scale with.\footnote{The reader is referred to \cite{mihai:plate} for an alternative theory for nematic elastomer plates.} 

\begin{figure}[h]
	\centering
	\includegraphics[width=.5\linewidth]{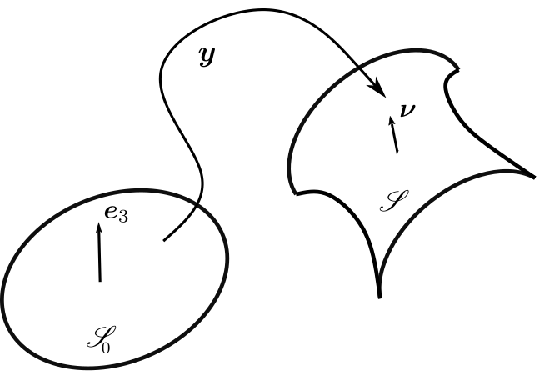}
	\caption{Schematic representation of the deformation of a thin sheet. $\surface_0$ is the planar midsurface of the slab $\slab$ of thickness $2h$ representing the reference configuration of the body. $\surface$ is the midsurface of the deformed slab $\f(\slab)$. $\surface_0$ lies in the plane $(\e_1,\e_2)$ of Cartesian frame $\cframe$. $\surface$ is the image under the mapping $\y$ of $\surface_0$; it is an oriented surface with unit normal $\normal$.}
	\label{fig:dimension_reduction}
\end{figure}

We identify the undeformed body with a slab $\slab$ of thickness $2h$ and midsurface $\surface_0$ in the $(\e_1,\e_2)$ plane of a Cartesian frame. We further assume that $\m$ is imprinted in $\slab$ so that it does not depend on the $x_3$ coordinate and  $\m\cdot\e_3\equiv0$. Moreover, we represent the three-dimensional deformation $\f$ as
\begin{equation}
\label{eq:deformation_representation}
\f(\x,x_3)=\y(\x)+\Phi(\x,x_3)\normal,
\end{equation} 
where $\x$ varies in $\surface_0$, $x_3$ ranges in the interval $[-h,h]$, $\normal$ is the normal to the midsurface $\surface=\y(\surface_0)$ of the deformed slab $\f(\slab)$ (see Fig.~\ref{fig:dimension_reduction}),
and $\Phi(\x,x_3)$ is a function to be determined,\footnote{In the classical theory of plates, the \emph{Kirchhoff-Love} hypothesis stipulates that $\Phi\equiv x_3$ (see, for example, \cite{podio:exact}, for a modern treatment). In \cite{ozenda:kirchhoff}, the Kirchhoff-Love hypothesis was reformulated in the more general form adopted here and criteria were suggested to identify the function  $\Phi$, none of which delivered exactly the original Kirchhoff-Love form.} enjoying the property
\begin{equation}\label{eq:Phi_property}
\Phi(\x,0)=0.	
\end{equation}	

As shown in \cite{ozenda:blend}, the constraint of incompressibility \eqref{eq:incompressibility} determines $\Phi$ in the form
\begin{equation}\label{eq:Phi_determination}
\Phi(\x,x_3)=x_3-Hx_3^2+\frac13(6H^2-K)x_3^3 +O(x_3^4),
\end{equation}
where $H$ and $K$ are the \emph{mean} and \emph{Gaussian} curvature of $\surface$, respectively, defined as
\begin{equation}
	\label{eq:H_and_K}
	H(\y(\x)):=\frac12\tr\curvature\quad\text{and}\quad K(\y(\x)):=\det\curvature,
\end{equation}
in terms of the two-dimensional curvature tensor $\nablas\normal$ at the point $\y(\x)$ on $\surface$.
The following formula for $\F$ was justified in \cite{ozenda:blend} as a consequence of \eqref{eq:Phi_property} for $h$ sufficiently small,
\begin{equation}
	\label{eq:F_formula}
	\F=\nabla\y+\Phi\nabla\normal+\Phi'\normal\otimes\e_3,
\end{equation}
where $\Phi'$ denotes the derivative of $\Phi$ with respect to $x_3$ and $\nabla$ is the gradient in $\x$, so that, in particular,
\begin{equation}
	\label{eq:nabla_normal}
	\nabla\normal=\curvature\nay.
\end{equation}

Since $\m\cdot\e_3\equiv0$, it follows from \eqref{eq:F_formula} and \eqref{eq:n_conveyed} that the nematic director $\n$ in the present configuration $\f(\slab)$ is delivered by
\begin{equation}
	\label{eq:n_present_configuration}
	\n(\f(\x,x_3))=\frac{(\I+\Phi\nablas\normal)\nay\m}{\lvert(\I+\Phi\nablas\normal)\nay\m\rvert},
\end{equation}
where \eqref{eq:nabla_normal} has also been used. Since $\nay\m\cdot\normal\equiv0$ and $\nablas\normal$ is a symmetric tensor mapping the local tangent plane to $\surface$ into itself, \eqref{eq:n_present_configuration} shows that $\n\cdot\normal=0$ everywhere within $\f(\slab)$, but $\n$ is \emph{not} uniform on the fibers along $\normal$, as $\Phi$ depends on $x_3$.

Moreover, by \eqref{eq:F_formula}, the three-dimensional right Cauchy-Green tensor $\Cf$ can be written as
\begin{equation}
	\label{eq:C_f_formula}
	\Cf=\C_\Phi+\Phi'^2\e_3\otimes\e_3,
\end{equation}
where
\begin{equation}
	\label{eq:C_Phi}
	\C_\Phi:=\C+\Phi\C_1+\Phi^2\C_2.
\end{equation}
In \eqref{eq:C_Phi}, 
\begin{equation}
	\label{eq:C_definition}
	\C:=\nay\trans\nay
\end{equation}
is the two-dimensional right Cauchy-Green tensor, 
\begin{equation}
	\label{C_1__C_2}
	\C_1:=2\nay\trans\curvature\nay,\quad\text{and}\quad\C_2:=\nay\trans\curvature^2\nay.
\end{equation}
By \eqref{eq:Phi_determination}, \eqref{eq:C_f_formula} becomes a (rather involved) function of the mapping $\y$ and the variable $x_3$.

Our next task is to integrate in the latter variable the expression for $f_\mathrm{t}$ in \eqref{eq:total_energy_density}. To this end, we make further use of the \emph{photo-uniformity}  approximation discussed at the end of Sect.~\ref{sec:model}: we shall take $(\lcper,\lcpar)$ and $S$ as \emph{independent} of $x_3$. In particular, by \eqref{eq:n_present_configuration} and \eqref{eq:Phi_property}, in \eqref{eq:phi_equation_unpolarized} we shall express $\n$ as
\begin{equation}
	\label{eq:n_photouniformity}
	\n(\f(\x,0))=\frac{\nay\m}{\lvert\nay\m\rvert}=\frac{\nay\m}{(\m\cdot\C\m)^{1/2}},
\end{equation} 
which is the nematic director evaluated on $\surface$ and appears to be conveyed by the deformation of $\surface_0$, in accordance with the three-dimensional constraint \eqref{eq:n_conveyed}.
Accordingly, to mimic \eqref{eq:incompressibility}, we shall also assume that  
\begin{equation}
	\label{eq:inextensibility}
	\det\C=1,
\end{equation}
a constraint that predicates the \emph{inextensibility} of the midsurface $\surface_0$.
\begin{comment}
Finally, since $\m\cdot\e_3\equiv0$, by \eqref{eq:n_conveyed}, \eqref{eq:F_formula}, and \eqref{eq:Phi_property}, the field $\n$ in \eqref{eq:n_photouniformity} equals for all $x_3\in[-h,h]$ the director evaluated at $\y(\x)\in\surface$ (i.e., for $x_3=0$).
\end{comment}

Reasoning as in \cite{ozenda:blend}, we identify the stretching and bending \emph{contents} of the surface 
\emph{elastic} free energy, as the $\y$-dependent components scaling like $h$ and $h^3$, respectively, of the three-dimensional density $f_\mathrm{t}$ integrated in $x_3$ over the thickness of $\slab$. Denoting by $F_\mathrm{s}$ and $F_\mathrm{b}$ the resulting stretching and bending elastic contents, respectively, both \emph{scaled} to $n_\mathrm{n}k_BTh$, with essentially the same computations illustrated in \cite{ozenda:blend}, which would be too boring to reproduce here, we arrive at
\begin{equation}\label{eq:F_s_formula}
\begin{split}
F_\mathrm{s}[S,\y{;}I]:=\mu\int_{\surface_0}\bigg\{&\frac{\lrper}{\lcper}+\frac{\lrper}{\lcpar}\tr\C+\frac{1}{\lcpar}(\lrpar-\lrper)\m\cdot\C\m\\&-\lrper\left(\frac{1}{\lcpar}-\frac{1}{\lcper}\right)\frac{1}{\m\cdot\C\m}\bigg\}\dd A
\end{split}
\end{equation}
and
\begin{equation}\label{eq:F_b_formula}
\begin{split}
F_\mathrm{b}[S,\y{;}I]:=\frac13h^2\mu\int_{\surface_0}\bigg\{&2\frac{\lrper}{\lcper}(8H^2-K)\\
&-\bigg[\frac{\lrper}{\lcpar}\tr\C+\frac{1}{\lcpar}(\lrpar-\lrper)\m\cdot\C\m\\
&+3\lrper\left(\frac{1}{\lcpar}-\frac{1}{\lcper}\right)\frac{1}{\m\cdot\C\m}\bigg]K\\
&+4\lrper\left(\frac{1}{\lcpar}-\frac{1}{\lcper}\right)(2H-\kappa_n)\kappa_n\frac{1}{\m\cdot\C\m}\bigg\}\dd A,
\end{split}
\end{equation}
where $A$ here denotes the area measure and
\begin{equation}
	\label{eq:kappa_n_definition}
	\kappa_n:=\n\cdot\curvature\n.
\end{equation}

$F_\mathrm{s}$ and $F_\mathrm{b}$ embody two separate components of the elastic free energy at the level of approximation we consider. To obtain the total free  energy $F_\mathrm{t}$ (likewise scaled to $n_\mathrm{n}k_BTh$), we must supplement $F_\mathrm{s}+F_\mathrm{b}$ with the integral (again across the slab's thickness) of the components of  $f_\mathrm{t}$ independent of the deformation $\y$,
\begin{equation}\label{eq:F_t_formula}
\begin{split}
F_\mathrm{t}[S,\y{;}I]&:=F_\mathrm{s}[S,\y{;}I]+F_\mathrm{b}[S,\y{;}I]\\
&+\int_{\surface_0}\bigg\{\mu\bigg[\ln\frac{\lcpar}{\lrpar}+2\ln\frac{\lcper}{\lrper}\bigg]+2(1-\phi)\psi_\mathrm{MS}(S)\bigg\}\dd A.
\end{split}
\end{equation}
It is worth noting that by the scaling chosen here  $F_\mathrm{t}$ has the physical dimensions of an area; to obtain a dimensionless energy, we should normalize $F_\mathrm{t}$ to the area of $\surface_0$ (which by \eqref{eq:inextensibility} is the same as the area of $\surface$), as will be done in the following section.

If light is impinging at right angles from above on $\surface_0$, so that $\kv=-\e_3$ (see Fig.~\ref{fig:dimension_reduction}), by \eqref{eq:n_photouniformity}, the \cs-population fraction $\phi$ in \eqref{eq:phi_equation_unpolarized} depends on $\kv$ and $\n$ through
\begin{equation}
	\label{eq:n_dot_k}
	(\n\cdot\kv)^2=\frac{(\e_3\cdot(\nabla\y)\m)^2}{\m\cdot\C\m}.
\end{equation}
\begin{comment}
We write explicitly $\F$ starting from (I15),
\begin{equation}
\label{eq:F}
\F=\nabla\y+\normal\otimes\e_3+x_3\nabla\normal,
\end{equation}	
from which it follows that
\begin{equation}
	\label{eq:C_f}
	\Cf=\C+\e_3\otimes\e_3+x_3\C_1+O(x_3^2),
\end{equation}
where
\begin{equation}
	\label{eq:C_1}
	\C_1:=[\nay\trans(\nabla\normal)+(\nabla\normal)\trans\nay].
\end{equation}
Now setting
\begin{equation}
	\label{eq:replacing}
	\tr\Cf=\tr\C+1+x_3\tr\C_1
\end{equation}
and integrating in $x_3$ over the interval $[-h,h]$, which makes the $x_3$-term disappear,
we arrive at the following formula for the  total free energy functional $F$ of a thin sheet (scaled to $n_\mathrm{n}k_BTh$),
\begin{equation}
	\label{eq:free_energy_functional}
	\begin{split}
	F[S,\y;\widetilde{I}]:=\int_{\surface_0}\bigg\{\widetilde{\mu}\bigg[&\frac{\lrper}{\lcper}+\frac{\lrper}{\lcpar}\tr\C+\frac{1}{\lcpar}(\lrpar-\lrper)\m\cdot\C\m\\&-\lrper\left(\frac{1}{\lcpar}-\frac{1}{\lcper}\right)\frac{1}{\m\cdot\C\m}
+\ln\frac{\lcpar}{\lrpar}\\&+2\ln\frac{\lcper}{\lrper}\bigg]+2(1-\phi)\psi_\mathrm{MS}(S)\bigg\}\dd A,
	\end{split}
\end{equation}
One of the advantages in adopting the same formalism of \cite{corbett:polarization} is that they provide a number of reasonable estimates for numerical parameters such as $a$, $b$, $A$, and $\widetilde{\mu}$, which can be readily used for our simulations.
\end{comment}

In \cite{corbett:polarization}, a few realistic values were suggested for the parameters that still need to be prescribed to solve a specific equilibrium problem. They suggested to take
\begin{equation}
	\label{eq:suggested_values}
	A=\frac16,\quad \frac{b}{a}=\frac12,\quad\text{and}\quad S_0\doteq0.61,\ \text{corresponding to}\ T\doteq0.91T_\mathrm{NI},
\end{equation}
where $T_\mathrm{NI}$ is the nematic-to-isotropic transition temperature. In our parameterization of the Maier-Saupe potential $\psi_\mathrm{MS}$ in \eqref{eq:Maier-Saupe_potential}, this value of $S_0$ corresponds to $J\doteq7.5$. As for the relative intensity $I$, Eisenbach~\cite{eisenbach:effect} suggests that it should not exceed $15$, whereas for Serra and Terentjev~\cite{serra:nonlinear} it could go up to $80$. In the application of our theory presented in following section, we shall take the above values for $A$, $b/a$, $S_0$, and $J$, and we shall never consider going beyond $I=70$.

\section{Ribbon Deflection}\label{sec:ribbon}
We now consider a thin, narrow ribbon originally parallel to the $(\bm{e}_1,\bm{e}_2)$ plane. Its midsurface
$\surface_0$ is a narrow strip of length $l$ and width $w$, represented by the set
$\surface_0=\{(x_1,x_2):0\leqq x_1\leqq l,\, 0\leqq x_2\leqq w\}$. We further assume that the ribbon
remains at all times homogeneous in $x_2$ and parallel to the $\bm{e}_2$ direction so that the scalar order parameter $S$ is
a function $S=S(x_1)$ of the $x_1$ coordinate only. 

We represent $\y$ as
\begin{equation}
	\label{eq:y_representation}
	\y(x_1,x_2)=y_1(x_1)\e_1+y_2(x_2)\e_2+y_3(x_1)\e_3,
\end{equation}
which allows the ribbon to come out of the $(\bm{e}_1,\bm{e}_2)$ plane while its normal remains
in the $(\bm{e}_1,\bm{e}_3)$ plane at all times. 

In order to make the inextensibility constraint \eqref{eq:inextensibility} explicit, we note that
\begin{equation}\label{eq:gradRibbon}
\nabla\y=y_1'(x_1)\bm{e}_1\otimes\bm{e}_1
+y_2'(x_2)\bm{e}_2\otimes\bm{e}_2+y_3'(x_1)\bm{e}_3\otimes\bm{e}_1
\end{equation}
and so
\begin{equation}
\C=\left\{[y_1'(x_1)]^2+[y_3'(x_1)]^2\right\}\bm{e}_1\otimes\bm{e}_1
+[y_2'(x_2)]^2\bm{e}_2\otimes\bm{e}_2.
\end{equation}
Hence, the requirement $\det\C=1$ becomes
\begin{equation}
\left\{[y_1'(x_1)]^2+[y_3'(x_1)]^2\right\}[y_2'(x_2)]^2=1. 
\end{equation}
Separation of variables shows that this is satisfied only if there are a scalar $\lambda>0$ and a function $\vt$ such that 
\begin{equation}\label{eq:CRibbon}
\C=\frac{1}{\lambda^2}\bm{e}_1\otimes\bm{e}_1
+\lambda^2\bm{e}_2\otimes\bm{e}_2
\end{equation}
and
\begin{equation}
	\label{eq:y_solution}
	y_1(x_1)=\frac{1}{\lambda}\int_0^{x_1}\cos\vt(x)\dd x, \quad 	y_3(x_1)=\frac{1}{\lambda}\int_0^{x_1}\sin\vt(x)\dd x,\quad y_2(x_2)=\lambda x_2.
\end{equation}
A value of $\lambda>1$ corresponds to a contraction of the length of the ribbon together with an extension of its width. 

The cross-section of the ribbon in the $(\bm{e}_1,\bm{e}_3)$ plane is described by the curve $\gamma$ where
\begin{equation}
\gamma(x_1)=\left(y_1(x_1),y_3(x_1)\right).
\end{equation}
As
\begin{equation}\label{eq:tangent}
\gamma'(x_1)=\left(\frac{1}{\lambda}\cos\vartheta(x_1),\frac{1}{\lambda}\sin\vartheta(x_1)\right),
\end{equation}
we see that $\vartheta(x_1)$ is simply the angle that the tangent to the ribbon makes at $\y(x_1,x_2)$ with the $\bm{e}_1$ axis.
A standard computation shows that there the curvature $\kappa$ of the ribbon is
\begin{equation}\label{eq:curvature}
\kappa(x_1)=\lambda \vert\vartheta'(x_1)\vert.
\end{equation}

\subsection{Ribbon free energy}
To make the free energy of the ribbon explicit, we write the director
imprinted in $\surface_0$ as
\begin{equation}
	\label{eq:n_0_representation}
	\m=\cos\vp_0\e_1+\sin\vp_0\e_2.
\end{equation}
As before, we assume that unpolarized light is shone upon the ribbon from above with $\kv=-\e_3$.
The right-hand-side of equation \eqref{eq:n_dot_k} can now be computed by using also
\eqref{eq:gradRibbon}, \eqref{eq:CRibbon}, and \eqref{eq:y_solution}; we find that
\begin{equation}
	\label{eq:n_dot_k_simplified}
	(\n\cdot\kv)^2=\frac{\sin^2\vt\cos^2\vp_0}{\cos^2\vp_0+\lambda^4\sin^2\vp_0}.
\end{equation}

At this point, it is worth remarking that the representation in \eqref{eq:y_representation} for $\y$ and hence the expression \eqref{eq:n_dot_k_simplified} are realistic only for either $\vp_0=0$ or $\vp_0=\frac\pi2$. For $0<\vp_0<\frac{\pi}{2}$, a more general class of deformations $\y$ should be considered, which also allows for twist.
As can be seen from \eqref{eq:n_dot_k_simplified}, the case $\vp_0=\frac\pi2$ is almost trivial, as the number fraction $\phi$ of \emph{cis} molecules given by equation \eqref{eq:phi_equation_unpolarized} would then be independent of $\vt$. Therefore, we focus on the case $\vp_0=0$, which will turn out to be rich enough to allow the ribbon to change shape.

The free energy density of the narrow ribbon is independent of $x_2$, and so the total free energy $F_\mathrm{t}$ given by \eqref{eq:F_t_formula}
can be simplified to
\begin{equation}\label{eq:free_energy_ribbon}
\begin{split}
F_\mathrm{t}[S,\lambda,\vt{;}I]:=\int_{0}^{l}\bigg\{\mu\bigg[&\frac43\lambda^2h^2\vt'^2+\frac{\lrper}{\lcper}+\frac{\lrper}{\lcpar}\left(\frac{1}{\lambda^2}+\lambda^2\right)\\
&+\frac{1}{\lcpar}(\lrpar-\lrper)\left(\frac{1}{\lambda^2}\cos^2\vp_0+\lambda^2\sin^2\vp_0\right)\\&-\lrper\left(\frac{1}{\lcpar}-\frac{1}{\lcper}\right)\frac{\lambda^2}{\cos^2\vp_0+\lambda^4\sin^2\vp_0}
\\&+\ln\frac{\lcpar}{\lrpar}+2\ln\frac{\lcper}{\lrper}\bigg]+2(1-\phi)\psi_\mathrm{MS}(S)\bigg\}\dd x_1,
\end{split}
\end{equation}	
where we have scaled to $n_\mathrm{n}k_BThw$ and only the integration over $x_1$ along the length of the ribbon remains.

Before introducing further simplifications, we give the free energy a dimensionless form by defining $\xi:=x_1/l$; the energy further
scaled to the length $l$ of the ribbon then becomes
\begin{equation}\label{eq:f_ribbon_dimless}
F_\mathrm{t}[S,\lambda,\vt{;}I]=\int_0^1\left\{\frac43\mu\left(\frac{h}{l}\right)^2\lambda^2\vt'^2+f(\vt,\lambda,S{;}I)\right\}\dd\xi,
\end{equation}
where 
\begin{equation}\label{eq:fOfTheta}
\begin{split}
f(\vt,\lambda,S{;}I):=\mu\bigg[&\frac{\lrper}{\lcper}+\frac{\lrper}{\lcpar}\left(\frac{1}{\lambda^2}+\lambda^2\right)
+\frac{1}{\lcpar}(\lrpar-\lrper)\left(\frac{1}{\lambda^2}\cos^2\vp_0+\lambda^2\sin^2\vp_0\right)\\&-\lrper\left(\frac{1}{\lcpar}-\frac{1}{\lcper}\right)\frac{\lambda^2}{\cos^2\vp_0+\lambda^4\sin^2\vp_0}+\ln\frac{\lcpar}{\lrpar}+2\ln\frac{\lcper}{\lrper}\bigg]
\\&+2(1-\phi)\psi_\mathrm{MS}(S)\,.
\end{split}
\end{equation}

\subsection{Clamped ribbon geometry}
We consider a ribbon \emph{clamped} at one end and \emph{free} at the other. Specifically,
we prescribe the boundary conditions
\begin{equation}\label{eq:boundaryConditions}
\vartheta(0)=0\quad\text{and}\quad\vartheta'(1)=0,
\end{equation}
that is, the curvature \eqref{eq:curvature} vanishes at the free end. We assume that the director is imprinted so as to
lie along the length of the ribbon, $\varphi_0=0$.

\begin{figure}[h]
\centering
\includegraphics[height=5cm]{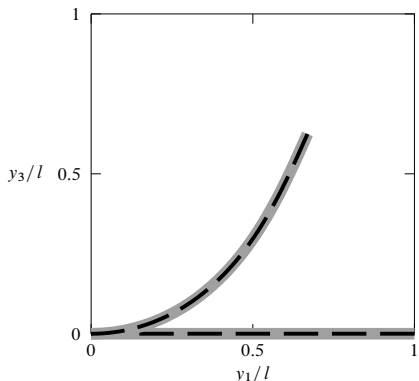}
\caption{The ribbon, depicted in grey, is clamped at the origin with its left end.  Without illumination, it lies
on the $y_1$-axis. Upon illumination, its right end is free to rise, with the curvature at the free end prescribed to be zero.
The black bars indicate the nematic director, which is aligned along the length of the ribbon.}\label{fig:Ribbon}
\end{figure}

By symmetry, $\vt\equiv0$ is always a critical point of $F_\mathrm{t}$. Indeed, $f$ in equation \eqref{eq:fOfTheta} is an \emph{even} function of $\vartheta$ for 
all values of the intensity $I$, see \eqref{eq:n_dot_k_simplified}. Therefore, all other critical points
of $F_\mathrm{t}$ come in \emph{pairs} of opposite signs, one member corresponding to an increasing function $y_3(\xi)$, the other to a decreasing function. We take the \emph{increasing} companion
as representative of each pair: $0\leqq\vt\leqq\vt_\mathrm{m}$, see the discussion at the end of Sect.~\ref{sec:model}. Figure~\ref{fig:Ribbon} shows a sketch of two ribbons, one undistorted and the other in a bent configuration.

\begin{figure}[h]
\includegraphics[height=5cm]{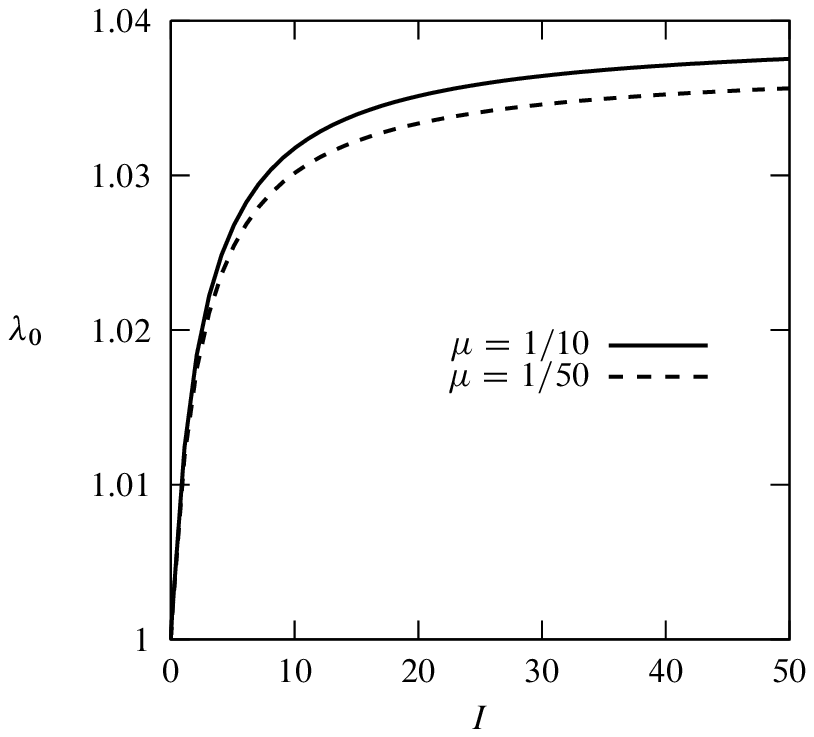}\hfill
\includegraphics[height=5cm]{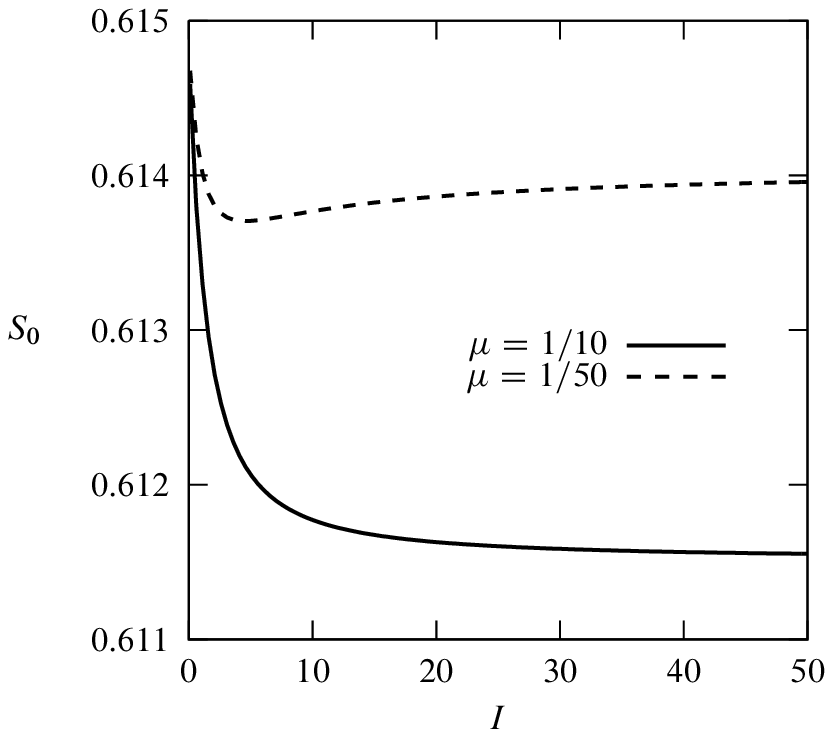}
\caption{Minimizers $(\lambda_0(I),S_0(I))$ of $f$ in \eqref{eq:fOfTheta} for the given light intensity
$I$ and $\vartheta=0$. These values also minimize the free energy \eqref{eq:f_ribbon_dimless} in the case $\vartheta\equiv0$.
Both the scalar order parameter $S$ and the stretching $\lambda$ vary on very slightly with the light intensity $I$.
The situation is similar for other values of $\vartheta$ (not shown).}\label{fig:SlambdavsI}
\end{figure}

The integrand in the free energy \eqref{eq:f_ribbon_dimless} has two distinct contributions:
the first term depends only on the derivative $\vartheta'$, and the second term, $f$ as given by \eqref{eq:fOfTheta}, depends only on $\vartheta$. If we assume that $\vartheta$ is constant, then the entire integrand is constant. For given light intensity $I$, the free energy is then
minimized if $f$ is the minimum with respect to $\lambda$ and $S$. In Fig.~\ref{fig:SlambdavsI} we show the minimising values
$\lambda_0$ and $S_0$ for the case $\vartheta\equiv0$ and the two values $\mu=1/10$ and $\mu=1/50$ for $0\le I\le 50$.

The magnitude of the deviation of $S$ and $\lambda$ from their values at zero light intensity remains very small for
all intensities. Motivated by this observation, we make the following simplification. We replace $f$ in \eqref{eq:fOfTheta}
by
\begin{equation}
	f_0(\vt{;}I):=f(\vt,\lambda_0(I),S_0(I){;}I)
\end{equation}
where $\lambda_0(I)$ and $S_0(I)$ are the minimizers for $f$ when $\vartheta=0$. The energy functional then becomes
\begin{equation}\label{eq:simpleF}
	F_\mathrm{t}[\vt{;}I]:=\int_0^1\left\{\frac43\mu\left(\frac{h}{l}\right)^2\left[\lambda_0(I)\right]^2\vt'^2+f_0(\vt{;}I)\right\}\dd\xi,
\end{equation}
which now depends on the single unknown function $\vartheta$.

\subsection{Euler-Lagrange equation}
Any equilibrium profile of the ribbon needs to satisfy the Euler-Lagrange equation derived from \eqref{eq:simpleF},
\begin{equation}\label{eq:EulerLagrange}
\frac{8}{3}\mu[\lambda_0(I)]^2\left(\frac{h}{l}\right)^2\vartheta''-\frac{\partial f_0(\vt{;}I)}{\partial\vartheta}=0.
\end{equation}
Since $\left.\frac{\partial f_0(\vt{;}I)}{\partial\vt}\right\vert_{\vt=0}=0$, the profile of the undistorted ribbon with 
$\vartheta\equiv 0$ is always
a solution of equation \eqref{eq:EulerLagrange} satisfying the boundary conditions \eqref{eq:boundaryConditions}; we shall call it the \emph{trivial} solution.
We analyze equation
\eqref{eq:EulerLagrange} using a hybrid approach: We first derive a range of explicit expressions that then serve as a basis for producing numerically a range of illustrating graphs using the parameters given in \eqref{eq:suggested_values}.

For non-trivial solutions, multiplying equation \eqref{eq:EulerLagrange} by $\vartheta'$ and integrating shows that a conservation law holds in the form
\begin{equation}\label{eq:conservation}
\frac{4}{3}\mu[\lambda_0(I)]^2\left(\frac{h}{l}\right)^2(\vartheta')^2-f_0(\vt{;}I)=c
\end{equation}
with a constant $c$. This constant can be determined by using the boundary condition $\vartheta'(1)=0$,
we have 
\begin{equation}\label{eq:constant}
c=-f_0(\vartheta(1){;}I).
\end{equation}
Equation \eqref{eq:conservation} is autonomous (i.e., it does not depend explicitly on $\xi$) and we are interested in its solutions with $\vartheta'\ge0$. On every such solution, if existing, $\vartheta$ ranges monotonically in the interval $[0,\vartheta(1)]$, and so we find that
$\vartheta(1)=\vartheta_\mathrm{m}$, namely that the largest ribbon angle is found at the free end. Moreover, it readily follows from \eqref{eq:tangent} that the corresponding equilibrium profile of the ribbon is \emph{convex} (as sketched in Fig.~\ref{fig:Ribbon}).

Using the constant \eqref{eq:constant} in equation \eqref{eq:conservation} we find that
\begin{equation}\label{eq:dtdx}
\vartheta'(\xi)=\frac{d\vartheta}{d\xi}=\frac{l}{2h\lambda_0(I)}\sqrt{\frac{3}{\mu}}\,\sqrt{f_0(\vartheta(\xi){;}I)-f_0(\vartheta_\mathrm{m}{;}I)},
\end{equation}
where for clarity we have here included the argument $\xi$. Separating variables and integrating yields
\begin{equation}
\xi=\frac{2h\lambda_0(I)}{l}\sqrt{\frac{\mu}{3}}\int_0^\vartheta\frac{\mathrm{d}\eta}{\sqrt{f_0(\eta{;}I)-f_0(\vartheta_\mathrm{m}{;}I)}}.
\end{equation}
Recalling that $\vartheta(1)=\vartheta_\mathrm{m}$, we find that
\begin{equation}\label{eq:Ffunction}
\frac{l}{h}=2\lambda_0(I)\sqrt{\frac{\mu}{3}}\int_0^{\vt_\mathrm{m}}\frac{\dd\vt}{\sqrt{f_0(\vt{;}I)-f_0(\vt_\mathrm{m}{;}I)}},
\end{equation}
which is a compatibility condition for the solution profile. It links the ratio $l/h$ of length to thickness of the ribbon,
the light intensity $I$, and the maximum angle $\vartheta_\mathrm{m}$.

\begin{figure}[h]
\begin{subfigure}[t]{0.48\linewidth}
	\centering
	\includegraphics[width=\textwidth]{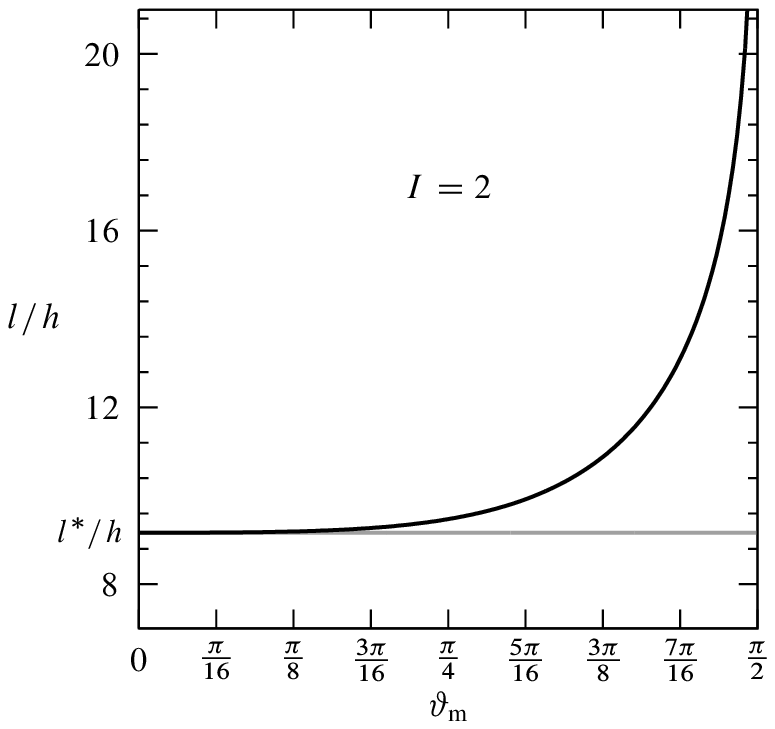}
	\caption{If $l/h$ is less than $l^\ast/h$, there is no value for $\vartheta_\mathrm{m}$ that allows for a bent profile, so the only possible equilibrium configuration is the trivial one. If $l/h$ is greater than $l^\ast/h$, there is a single solution with a bent profile.}
\end{subfigure}\hfill
\begin{subfigure}[t]{.48\linewidth}
	\centering
	\includegraphics[width=\textwidth]{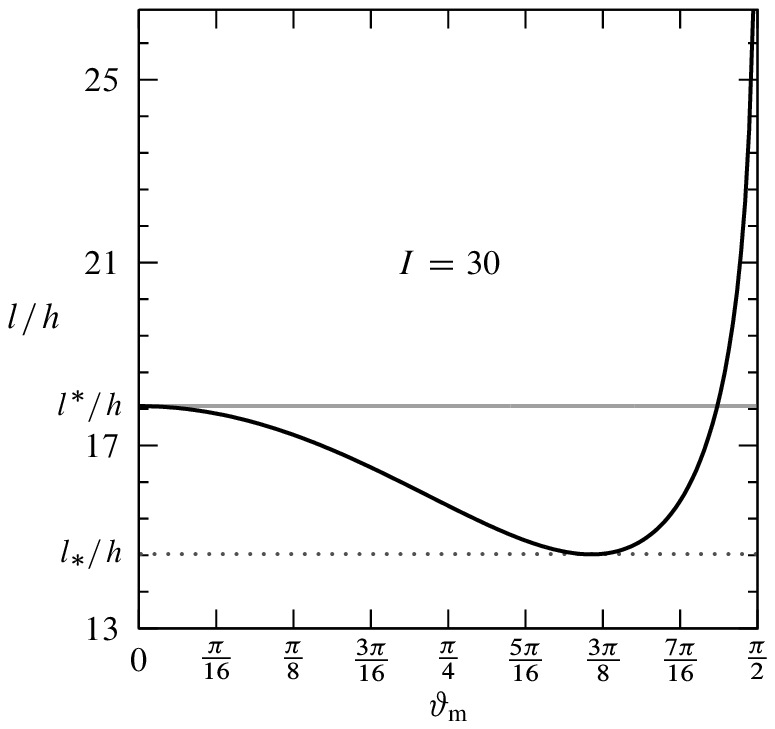}
	\caption{For $l/h$ less than the minimum $l_\ast/h $ of the black graph, only the trivial solution exists. For $l/h$ bigger than the minimum but less than $l^\ast/h$, two bent solutions exist. Finally, when $l/h>l^\ast/h$ there is only one bent solution.}
\end{subfigure}
\caption{For $I=2$ in (a) and $I=30$ in (b) the black graph shows the values of  $l/h$ required to satisfy the compatibility condition \eqref{eq:Ffunction} as a function of $\vartheta_\mathrm{m}$. The grey line shows the limit $l^\ast/h$ of $l/h$ as $\vartheta_\mathrm{m}\rightarrow 0$, see equation \eqref{eq:lstar}.
For $I=2$, there are two different regimes depending on the value of $l/h$, while for $I=30$, there are three regimes.}\label{fig:Fdata}
\end{figure}

A condition for a bent solution to bifurcate from the trivial one can be obtained by computing the
limit of equation \eqref{eq:Ffunction} as $\vartheta_\mathrm{m}\rightarrow 0$, assuming that it exists. To find the limit of the integral, we
consider the Taylor series of $f_0$ near zero,
\begin{equation}\label{eq:Taylor2}
 f_0(\vt)=f_0(0)+\frac{1}{2}f_0''(0)\vt^2+O(\vt^4),
\end{equation}
where, for brevity, the parameter $I$ has been dropped  from the argument of $f_0$ and a prime denotes differentiation with respect to $\vartheta$. Use of \eqref{eq:Taylor2} in \eqref{eq:Ffunction} shows that the existence of the limit requires
to have $f_0''(0)<0$, as
\begin{equation}
\lim_{\vartheta_\mathrm{m}\rightarrow 0}\int_0^{\vartheta_\mathrm{m}} \frac{\dd\vt}{\sqrt{f_0(\vt)-f_0(\vt_\mathrm{m})}}
=
\sqrt{\frac{-2}{f_0''(0)}}\lim_{\vartheta_\mathrm{m}\rightarrow 0}\int_0^{\vartheta_\mathrm{m}} \frac{\dd\vt}{\sqrt{\vt_\mathrm{m}^2-\vt^2}}
=\frac{\pi}{2}\sqrt{\frac{-2}{f_0''(0)}},
\end{equation}
and so a bifurcation is found when
\begin{equation}\label{eq:lstar}
\frac{l^\ast}{h}=\pi\lambda_0(I)\sqrt{\frac{-2\mu}{3f_0''(0)}}. 
\end{equation}

Figure~\ref{fig:Fdata} shows for the two intensities $I=2$ and $I=30$ the value of $l^\ast/h$ as a grey horizontal line and, as a black graph, the values of $l/h$ obtained using the compatibility condition \eqref{eq:Ffunction} for $0<\vartheta_\mathrm{m}<\frac{\pi}{2}$. The two pictures are qualitatively very different. For the lower intensity, there is a single critical value $l^\ast/h$ below which no bent solution is possible, and above which a single bent solution
is present. For the higher intensity, in addition to $l^\ast/h$, there is a second critical value $l_\ast/h$ below which no bent solution exist. For 
$l_\ast/h<l/h<l^\ast/h$ there are two bent solutions, and for $l/h>l^\ast/h$ there
is a single bent solution, as in the other case.

To determine the critical value $I_\mathrm{c}$ of the intensity at which the two different behaviours meet, we examine the compatibility condition
\eqref{eq:Ffunction} for small values of $\vartheta_\mathrm{m}$. To this end, we need one further term of the Taylor series of
$f_0$ near zero,
\begin{equation}\label{eq:Taylor4}
 f_0(\vt)=f_0(0)+\frac{1}{2}f_0''(0)\vt^2+\frac{1}{24}f_0^{(iv)}(0)\vt^4+O(\vt^6).
\end{equation}
A straightforward computation then shows that
\begin{equation}
 \int_0^{\vartheta_\mathrm{m}} \frac{\dd\vt}{\sqrt{f_0(\vt)-f_0(\vt_\mathrm{m})}}
=\frac{\pi}{2}\sqrt{\frac{-2}{f_0''(0)}}\left(1-\frac{1}{16}\vartheta_\mathrm{m}^2\frac{f_0^{(iv)}(0)}{f_0''(0)}\right)+O(\vartheta_\mathrm{m}^4).
\end{equation}
Therefore, given that $-f_0''(0)$ is positive, when $f_0^{(iv)}(0)$ is
positive we have the situation shown in Fig.~\ref{fig:Fdata} (a) with a
single critical value $l^\ast/h$. When $f_0^{(iv)}(0)$ is negative
we have the situation shown in Fig.~\ref{fig:Fdata} (b) with the additional
critical value $l_\ast/h$.

\begin{figure}[h]
\includegraphics[width=0.3\textwidth]{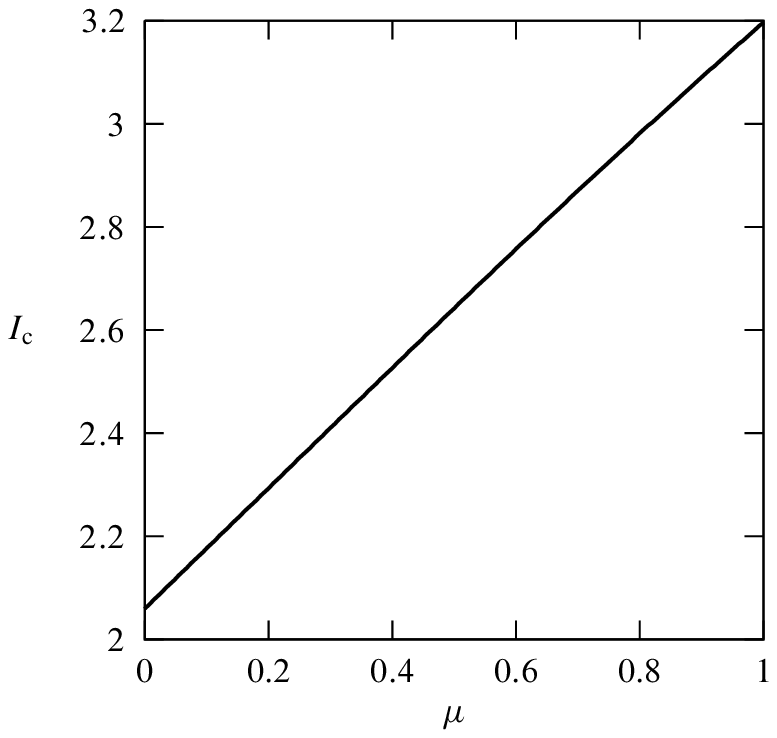}\hfill
\includegraphics[width=0.3\textwidth]{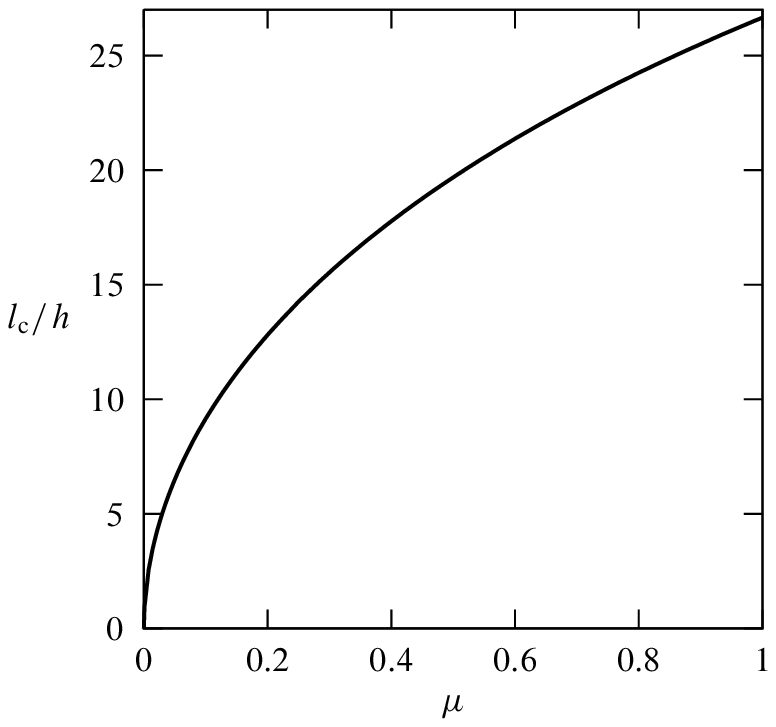}\hfill
\includegraphics[width=0.3\textwidth]{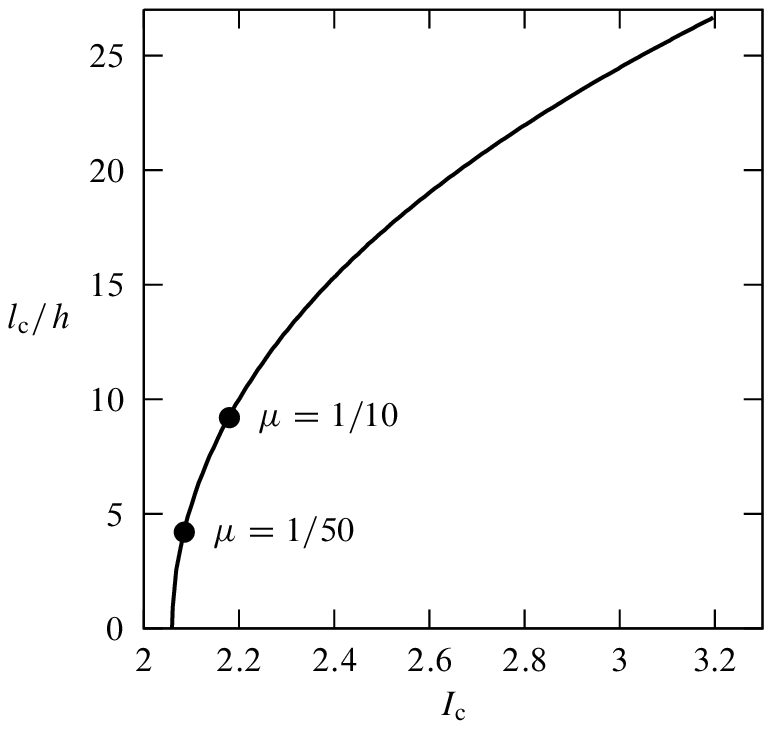}
\caption{Critical intensity $I_\mathrm{c}$ (left) and critical length $l_\mathrm{c}/h$ (middle) versus $\mu$ and
the curve $(I_\mathrm{c},l_\mathrm{c}/h)$ parameterized by $\mu$ (right) with the relevant values of $\mu$ marked by black circles. For length-to-thickness ratios
below the graph in the middle diagram, no bent solutions exist for any intensity.}\label{fig:criticalVsMu}
\end{figure}

On the left of Fig.~\ref{fig:criticalVsMu} we show for the whole possible range $0\le\mu\le 1$ the critical values $I_\mathrm{c}$ of the intensity where the fourth dervative of $f_0$ at zero changes sign and in the middle the corresponding critical values $l_\mathrm{c}/h$ obtained by equation \eqref{eq:lstar}, $l_\mathrm{c}$=$l^\ast(\mu,I_\mathrm{c})$. On the right, parametrized by $\mu$,
we plot the curve of critical points $(I_\mathrm{c},l_\mathrm{c}/h)$ in the $I$-$l/h$ plane.

\begin{figure}[b]
\begin{subfigure}[t]{0.48\linewidth}
	\centering
	\includegraphics[width=\textwidth]{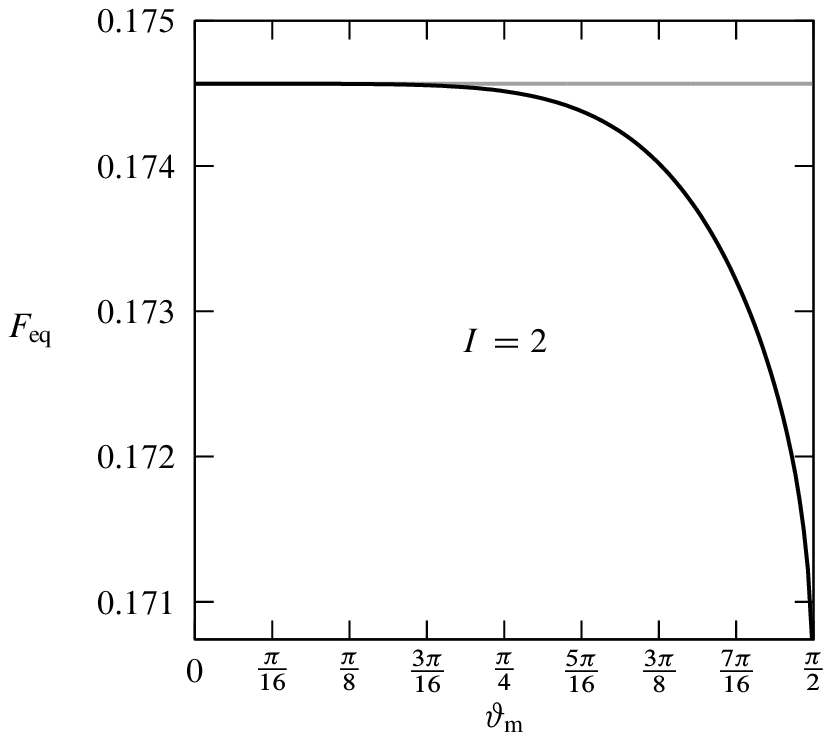}
	\caption{For $I=2$, if a bent solutions exists, it is always stable as it has lower energy
than the trivial solution.}
\end{subfigure}\hfill
\begin{subfigure}[t]{.48\linewidth}
	\centering
	\includegraphics[width=\textwidth]{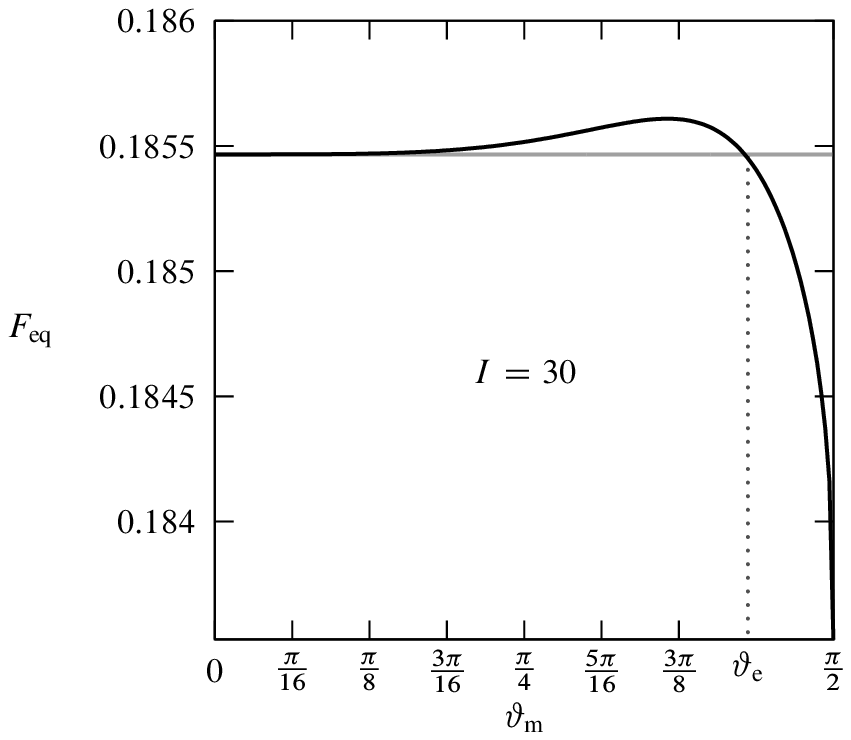}
	\caption{For $I=30$, both bent solutions with lower and with higher energy
than the trivial solution exist, compare Fig.~\ref{fig:Fdata} (b). There is exactly one bent solution with the same energy as the trivial solution; its corresponding angle $\vartheta_\mathrm{m}$ is marked in the figure as $\vartheta_\mathrm{e}$.}
\end{subfigure}
\caption{Energy of equilibrium solutions for $I=2$ and $I=30$ as a function of
$0\le\vartheta_\mathrm{m}\le\frac{\pi}{2}$. The length $l/h$ of the ribbon is implicitly determined by
the compatibility condition \eqref{eq:Ffunction} and can be read off from Fig.~\ref{fig:Fdata}. For reference, the grey line shows the energy of the undistored ribbon.}\label{fig:Edata}
\end{figure}

To assess the stability of bent ribbon profiles relative to the $\vartheta\equiv0$ profile, we compare energies. In this (limited) perspective we shall say that an equilibrium configuration is stable if it has less energy than any other configuration.

The total energy of the ribbon in an equilibrium configuration can be expressed as a function
of $\vartheta_\mathrm{m}$ by using in the functional \eqref{eq:simpleF} the conservation law \eqref{eq:conservation} with the constant from \eqref{eq:constant}:
\begin{align}
F_\mathrm{eq}(\vt_\mathrm{m}{;}I)&=\int_0^1\left\{f_0(\vt{;}I)-f_0(\vt_\mathrm{m}{;}I)+f_0(\vt{;}I)\right\}\dd\xi \nonumber\\
&=\int_0^{\vartheta_\mathrm{m}}\left\{\frac{2f_0(\vt{;}I)-f_0(\vt_\mathrm{m}{;}I)}{\sqrt{f_0(\vartheta{;}I)-f_0(\vartheta_\mathrm{m}{;}I)}}\frac{2h\lambda_0(I)}{l}\sqrt{\frac{\mu}{3}}\right\}\dd\vartheta\nonumber\\
&=\frac{1}{\int_0^{\vt_\mathrm{m}}\frac{\dd\vt}{\sqrt{f_0(\vt{;}I)-f_0(\vt_\mathrm{m},I)}}}\int_0^{\vt_\mathrm{m}}\frac{2f_0(\vt{;}I)-f_0(\vt_\mathrm{m}{;}I)}{\sqrt{f_0(\vt{;}I)-f_0(\vt_\mathrm{m}{;}I)}}\dd\vt,
\end{align}
where in the first step we have used equation \eqref{eq:dtdx} for substituting $\xi$ with $\vartheta$, and in the second step
we have used the condition \eqref{eq:Ffunction} to replace $l/h$.

Figure~\ref{fig:Edata} shows the energy of equilibrium solutions for the same two intensities used in
Fig.~\ref{fig:Fdata}, $I=2$ and $I=30$, for $0\le\vartheta_\mathrm{m}\le\frac{\pi}{2}$. The value
of $l/h$ is implicitly defined by the compatibility condition \eqref{eq:Ffunction}. For $I<I_\mathrm{c}$ (Fig.~\ref{fig:Edata} (a)), if
a bent solution exists it is always stable. For $I>I_\mathrm{c}$ (Fig.~\ref{fig:Edata} (b)), there can be both unstable and stable bent solutions; the maximum of $F_\mathrm{eq}$ is attained for the same value of $\vartheta_\mathrm{m}$ as the minimum $l_\ast/h$ of the function defined by \eqref{eq:Ffunction} (see Fig.~\ref{fig:Fdata} (b)). As $l/h$ is increased above $l_\ast/h$ two bent solutions originate whose energies fall on either side of the maximum in Fig.~\ref{fig:Edata} (b). As $l/h$ keeps increasing, the solution with the smaller $\vartheta_\mathrm{m}$ gradually approaches the trivial solution,  having always higher energy until it merges with it at $l/h=l^\ast/h$. Correspondingly, the solution with larger $\vartheta_\mathrm{m}$ keeps reducing its energy: there is then exactly one bent solution with the same energy as the trivial solution; we
denote its maximum deflection angle $\vartheta_\mathrm{m}$ by $\vartheta_\mathrm{e}$ and the corresponding length to thickness ratio
by $l_\mathrm{e}/h$.

\subsection{Bifurcation analysis}
\begin{figure}[h]
\begin{center}
\includegraphics[width=0.48\textwidth]{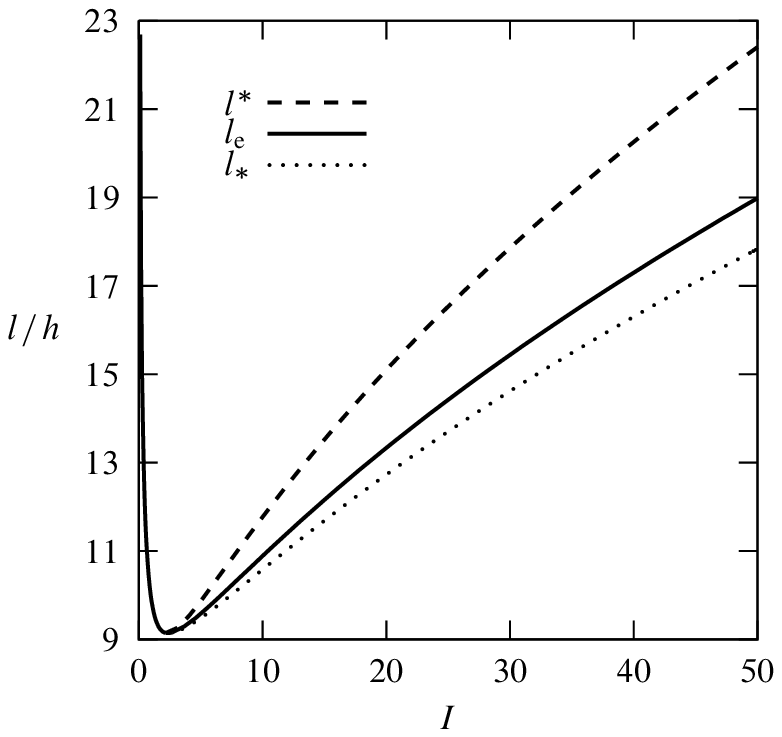}\hfill\includegraphics[width=0.48\textwidth]{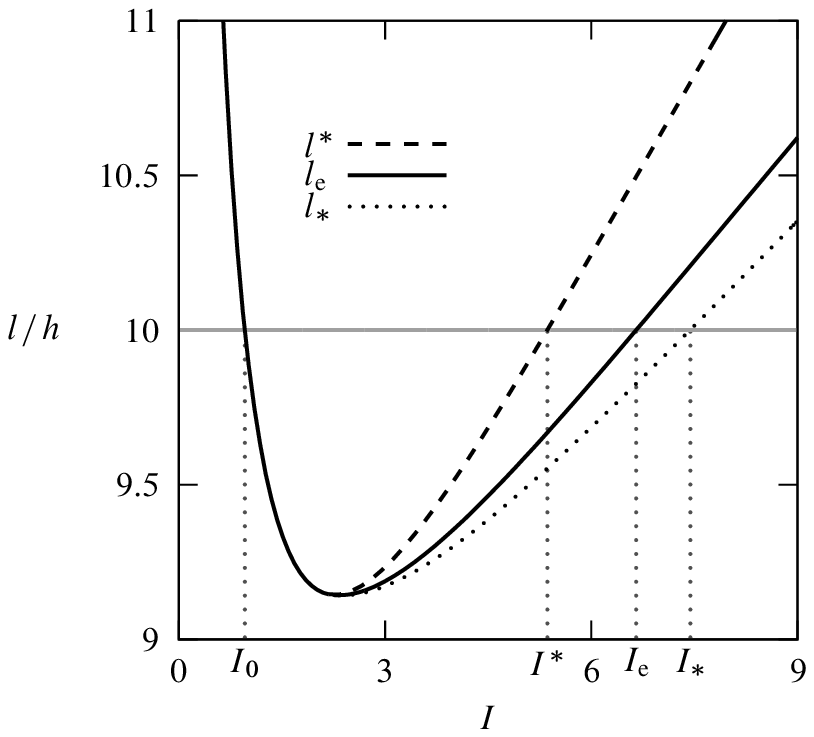} 
\end{center}
\caption{Critical lengths versus light intensity for $\mu=1/10$. Along the solid line, a bent solution exists that has the same energy as the trivial solution. Above the dashed line, the trivial solution is unstable. Below the dotted line, no bent solution exists. To the left of the minimum, located at the critical point $(I_\mathrm{c},l_\mathrm{c}/l)$, all three critical lengths coincide and only the solid line is drawn: below it no bent solution exists, above it the trivial solution is unstable. In the blow-up on the right the grey line marks $l/h=10$, and its intersections with the graphs are marked $I_0$, $I^\ast$, $I_\mathrm{e}$, and  $I_\ast$: the same intensities are also identified in Fig.~\ref{fig:Bifurcation10} below.}\label{fig:LvsI}
\end{figure}

To summarize what we have established so far, we present in Fig.~\ref{fig:LvsI}
three graphs in the $I$-$l/h$ plane: the solid line shows $l_\mathrm{e}/h$, where a bent solution exists that has the same energy as the trivial solution. The dashed line shows $l^\ast/h$ as given by \eqref{eq:lstar}: above this line the trivial solution is unstable. The dotted line corresponds to the minimum $l_\ast/h$ of the graph shown in Fig.~\ref{fig:Fdata}~(b): below this line, no bent solution exists.
The minimum of the graph in Fig.~\ref{fig:LvsI} has coordinates $(I_\mathrm{c},l_\mathrm{c}/h)$ for the chosen value of $\mu=1/10$, see the right graph in Fig.~\ref{fig:criticalVsMu}.

\begin{figure}[h]
\begin{center}
\includegraphics[width=0.9\textwidth]{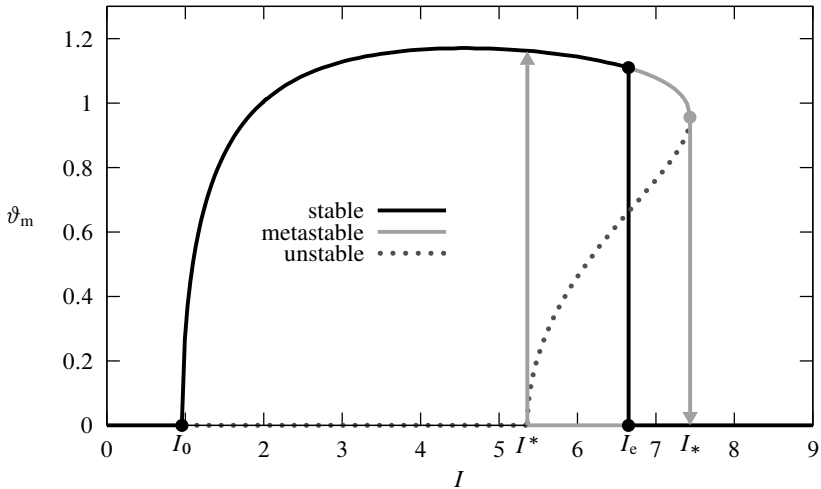} 
\end{center}
\caption{Bifurcation diagram for $l/h=10$. The intensities marked $I_0$, $I^\ast$, $I_\mathrm{e}$, and  $I_\ast$ correspond to those
shown in the right graph in Fig.~\ref{fig:LvsI}. As the intensity is increased from zero, at $I_0$ the trivial solution becomes unstable and a bent solution bifurcates up from $\vartheta=0$. At $I^\ast$, the trivial solution enters the scene again, but here has higher energy than the bent solution. Both solutions have equal energy at $I_\mathrm{e}$, and for intensities above $I_\ast$ no bent solution exists any more.}\label{fig:Bifurcation10}
\end{figure}

While Fig.~\ref{fig:LvsI} contains the gist of our analysis, it is rather artificial in that
it shows length-to-thickness ratios as functions of the light intensity. Of course, in an experiment
$l/h$ is fixed and only $I$ can be varied. We display therefore in Fig.~\ref{fig:Bifurcation10}
the maximum angle $\vartheta_\mathrm{m}$ of the ribbon profile as a function of the intensity. When the light intensity reaches $I_0$, a bifurcation
from the trivial to a bent solution occurs. With increasing intensity, the maximum angle $\vartheta_\mathrm{m}$ first increases but eventually decreases, and at $I_\ast$ the bent solution disappears altogether.
Upon then decreasing the light intensity, the trivial solution can be continued up to $I^\ast$, where it merges with the unstable bent solution. In the interval between $I^\ast$ and $I_\ast$ there are \emph{three} equilibria, the trivial solution and \emph{two} bent solutions; our elementary stability taxonomy is not sufficient to cover all cases: we then call \emph{metastable} the solution with intermediate energy, and stable (as before) the one with the least energy (see Fig.~\ref{fig:Bifurcation10}). One bent solution is stable in $(I^\ast,I_\mathrm{e})$ and the trivial solution is metastable, whereas the trivial solution is stable in  $(I_\mathrm{e},I_\ast)$ and the same bent solution is metastable. In both intervals one bent solution is always unstable, the one that merges with the trivial solution at $I^\ast$. The intensities $I^\ast$ and $I_\ast$ delimit a \emph{hysteresis} loop, which encloses the intensity $I_\mathrm{e}$ where the stable bent solution has the same energy as the trivial solution; there a first-order \emph{shape transition} takes place, which could be seen as an abrupt \emph{snapping} back and forth of the ribbon.

\begin{figure}[h]
\begin{center}
\includegraphics[width=0.9\textwidth]{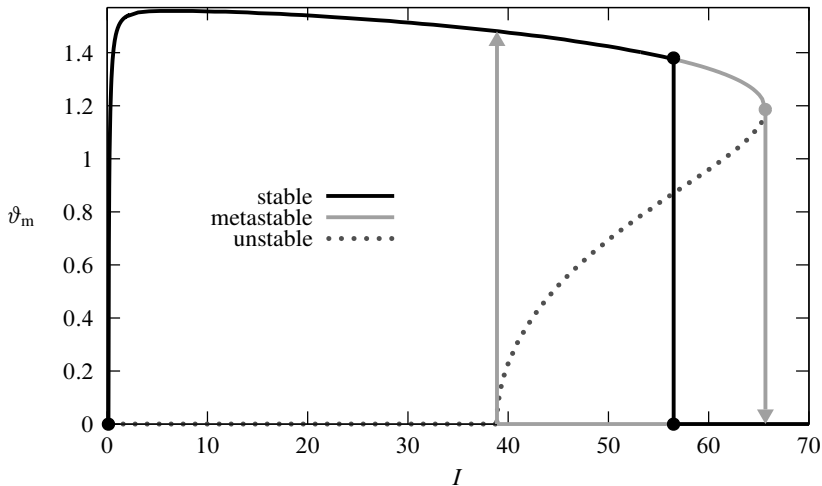} 
\end{center}
\caption{Bifurcation diagram for $l/h=20$. The structure is the same as in Fig.~\ref{fig:Bifurcation10}. The relatively longer ribbon here leads to a bent solution that both occurs at a lower intensity and is sustained for higher intensities. Also, the maximum angle is closer to $\pi/2$.}\label{fig:Bifurcation20}
\end{figure}

The picture is similar for all values of $l/h>l_\mathrm{c}/h$. As an example, we show in 
Fig.~\ref{fig:Bifurcation20} the bifurcation diagram for $l/h=20$. For $l/h<l_\mathrm{c}/h$, there
is literally nothing to see: no bent solutions exist for any value of the intensity.

\begin{figure}[h]
\includegraphics[height=2.7cm]{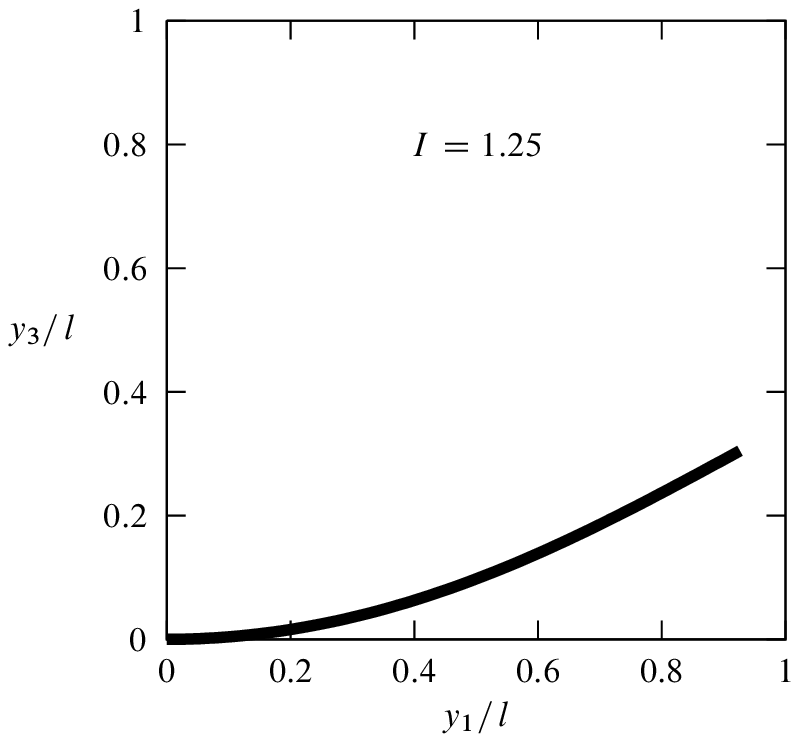}\hfill%
\includegraphics[height=2.7cm]{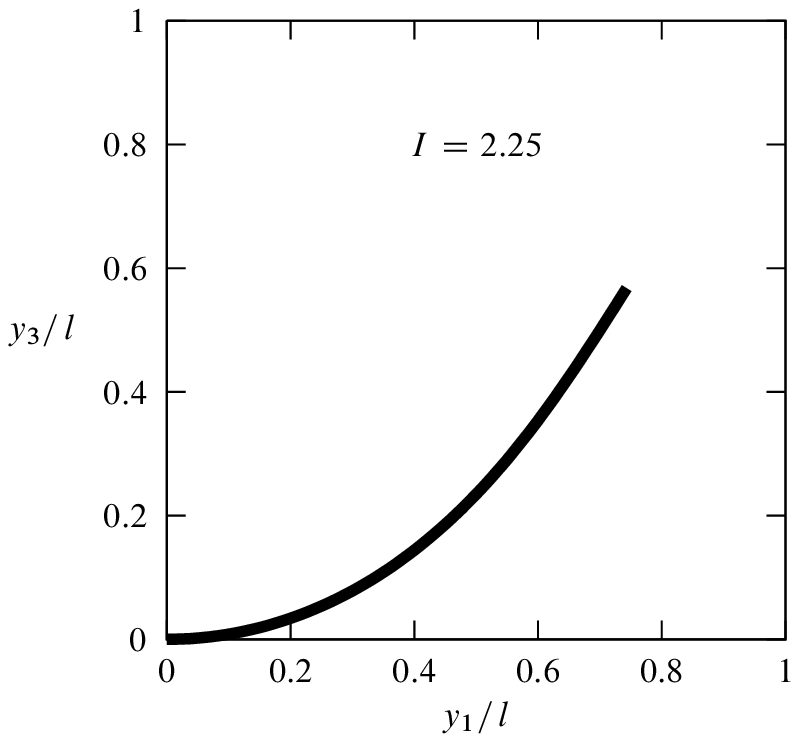}\hfill%
\includegraphics[height=2.7cm]{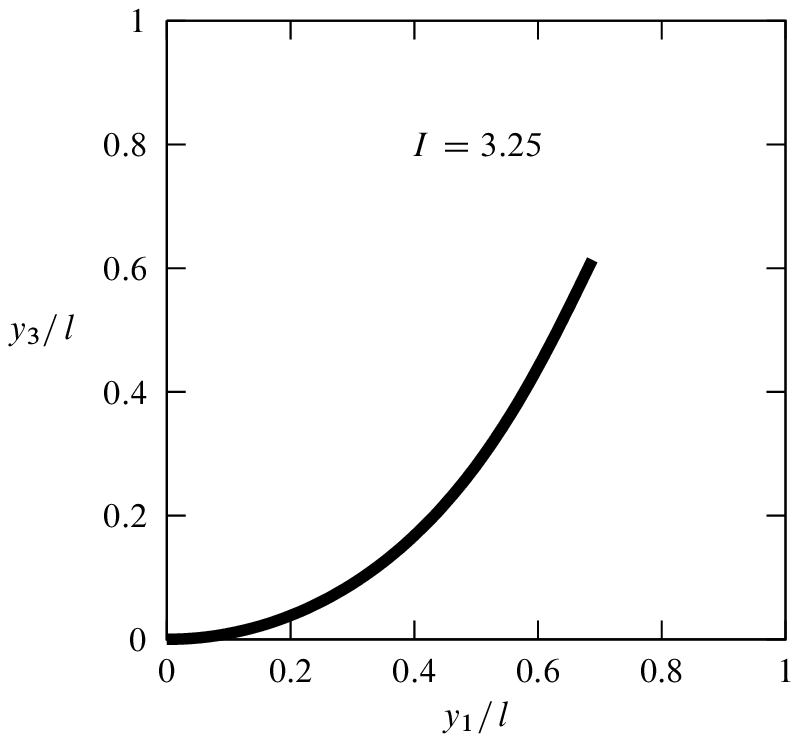}\hfill%
\includegraphics[height=2.7cm]{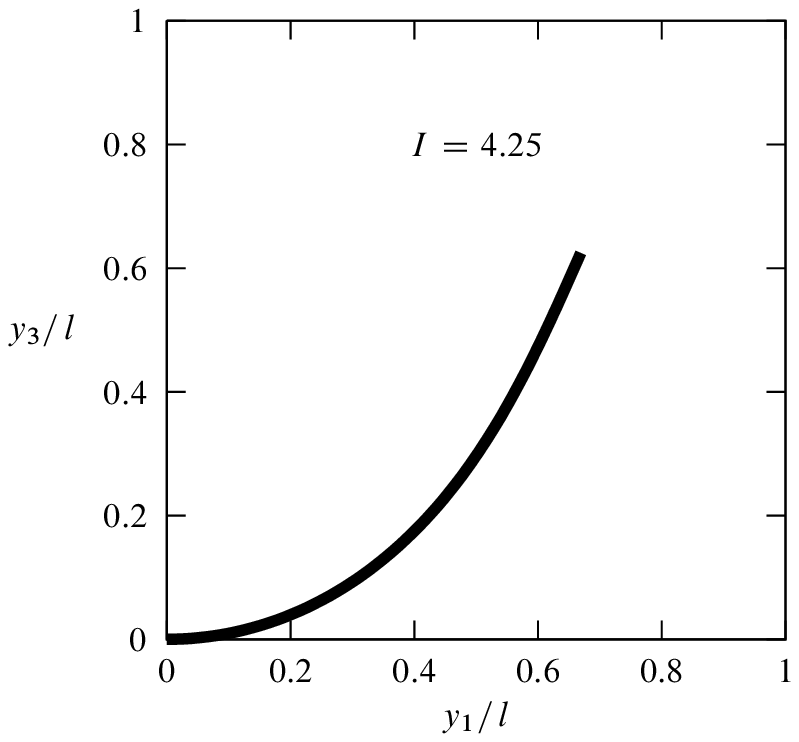}\\
\includegraphics[height=2.7cm]{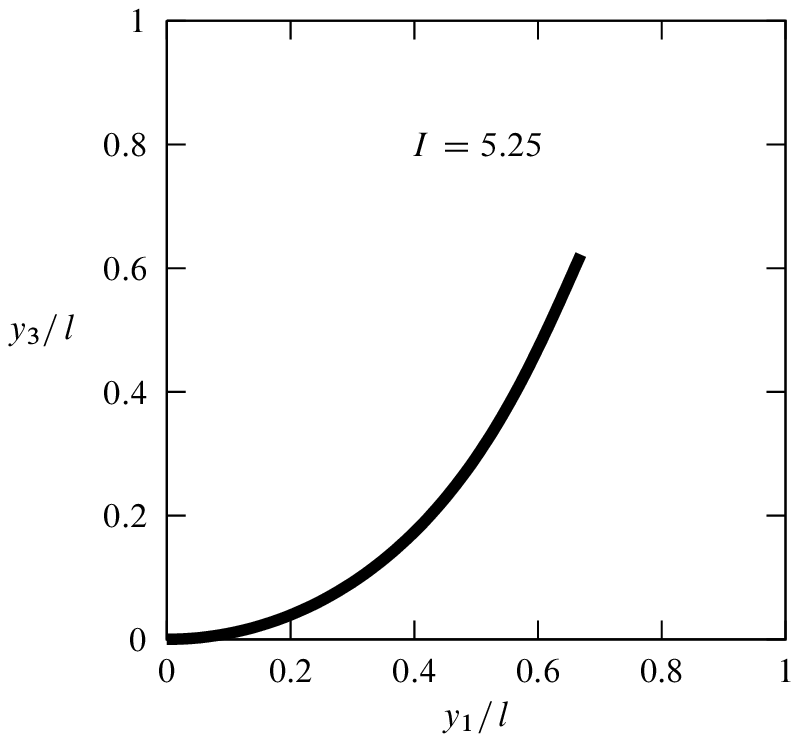}\hfill%
\includegraphics[height=2.7cm]{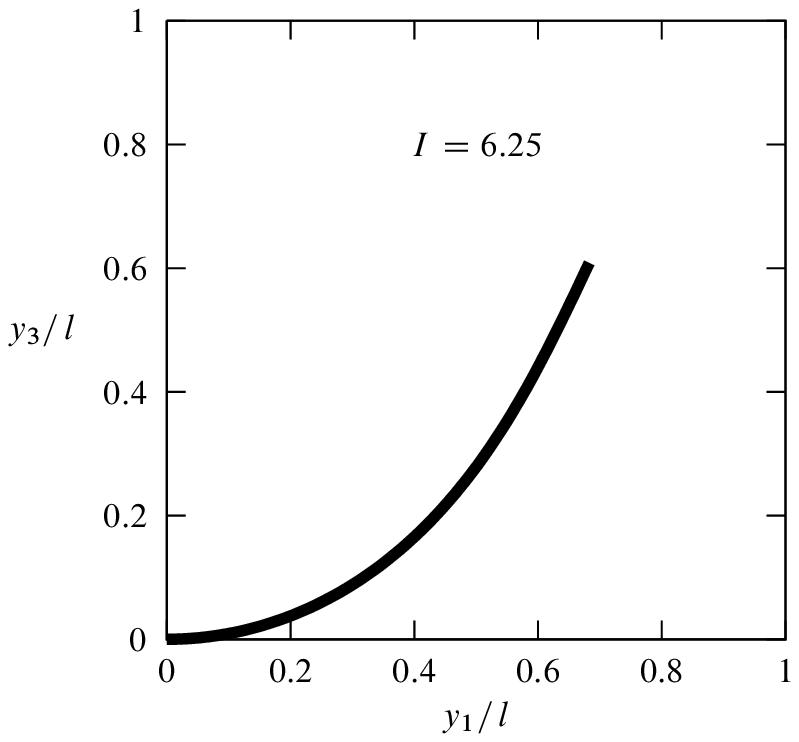}\hfill%
\includegraphics[height=2.7cm]{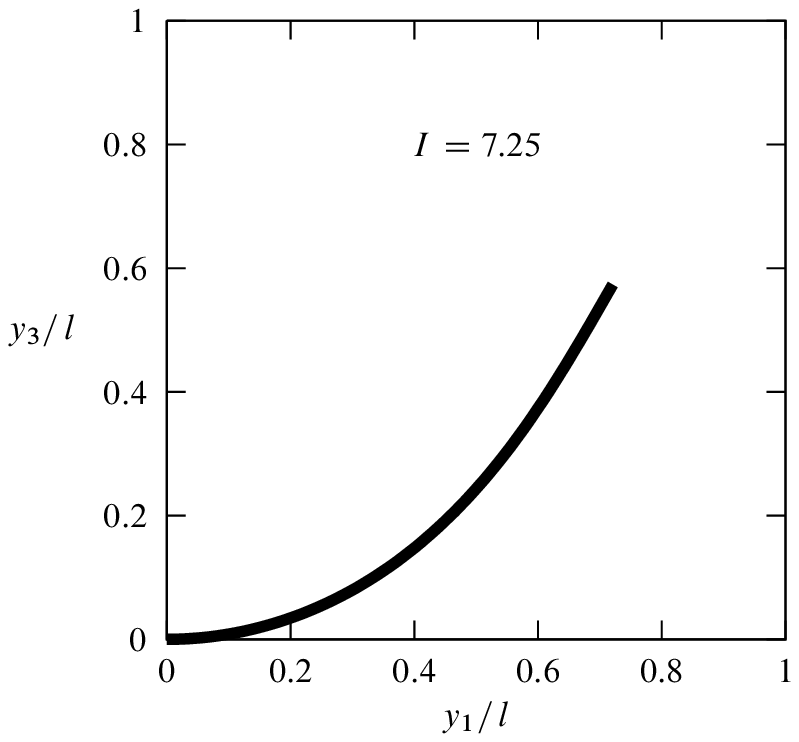}\hfill%
\includegraphics[height=2.7cm]{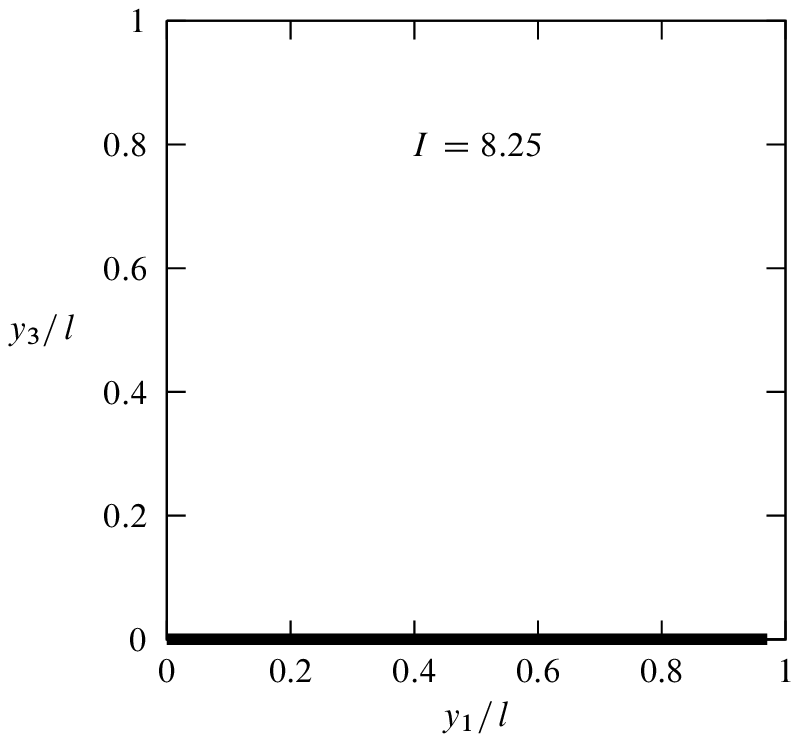}
\caption{Ribbon profiles for varying intensities with $\mu=1/10$ and $l/h=10$. Without illumination, the ribbon occupies the entire space on the $y_1$ axis between zero and one. When the illumination intensity exceeds a threshold value, the ribbon begins to extend upwards. With increasing intensity, the ribbon first extends further up before reaching a maximum height. For even higher extensities, the height first decreases before the ribbon eventually snaps back to the $y_1$ axis. In the final picture shown, the ribbon is back on the $y_1$ axis but does not reach the point $y_1=1$ because for $I>0$ also $\lambda>1$.}\label{fig:Profiles}
\end{figure}

Finally, we show in Fig.~\ref{fig:Profiles} numerically computed solutions of the Euler-Lagrange equation \eqref{eq:EulerLagrange} satisfying the boundary
conditions \eqref{eq:boundaryConditions}. The parameters used are $\mu=1/10$ and $l/h=10$ for a range of intensities
between $I=1.25$ and $I=8.25$, as shown. The angles of the ribbon on its right end coincide with the values of $\vartheta_\mathrm{m}$ shown in Fig.~\ref{fig:Bifurcation10}.

\section{Conclusions}\label{sec:conclusion}
We presented a continuum model for photoresponsive elastomers, whose deformation is driven by illumination. The (dynamical) interaction of light with a nematogenic polymer network (hosting photoresponsive molecules) was described within the statistical mechanics model of Corbett and Warner \cite{corbett:linear,corbett:polarization}.

Light is responsible for a change in shape of photoresponsive molecules, which, once activated, deplete the nematic phase, and so have an effect on the nematic sclar order parameter $S$. This is what makes these materials different from ordinary nematic elastomers, where a change in $S$ is thermally induced. The spontaneous deformation that ensues photoactivation is likely to change illumination conditions, and this in turn may either enhance or hamper deformation. 

Our main objective was to study the equilibrium of a ribbon illuminated at right angles on one face in its undeformed configuration. To this end, we performed the  reduction of the total free-energy functional to a thin planar sheet and applied our reduced theory to a ribbon with a realistic choice of physical parameters collected from the specialized literature. We found the equilibrium configurations of the illuminated ribbon with the nematic director frozen along its longer side and represented in closed form their bifurcation scenario. 

We proved that the activation process is neither linear nor monotonic: the deflection of an activated ribbon first increases as expected, but then decreases before vanishing abruptly upon increasing the light intensity $I$ above a critical value $I_\mathrm{e}$, where the ribbon undergoes a first-order \emph{shape transition}. A hysteresis cycle is present about  $I=I_\mathrm{e}$, featuring two metastable configurations of the ribbon. Our bifurcation analysis has also shown that for a given length of the ribbon there is an optimal value of $I$ for which the deflection is maximum.

While deflections and displacements are generally large, dilations and contractions in an illuminated ribbon remain small. However, there is a critical length of the ribbon (depending only on the degree of cross-linking in the material) below which \emph{no} deflection can be promoted by light, no matter how intense this is: the system is too stiff to gain energy by bending.

Heuristically, the non-monotonic response of an illuminated ribbon, with its abrupt fall, could be explained by the coupling between the \cs-population fraction $\phi$ and the illumination angle that the direction of propagation of light $\kv$ makes with the nematic director $\n$ (tangent to the ribbon's longer side in our case). We remarked that for $S>0$ this coupling is most effective when $\kv$ and $\n$ are orthogonal; when the spontaneous deformation reduces the illumination angle, photoactivation is reduced, resulting in a negative feedback.

In more mundane terms, we may say that when activating a ribbon with light, we should be \emph{gentle}: too high an intensity may easily result in no deflection.
 
Our theory has limitations too. Perhaps the most conspicuous one is the \emph{photo-uniformity} approximation that has been used at various stages. Since we do not account for partial penetration of light in the ribbon's cross-section, deflection either towards light or away from it would have precisely the same energy cost. That deflection actually takes place \emph{towards} light (as experimentally observed) should be inferred from \emph{ad hoc} extrinsic considerations. Attempts have recently been made to  account for the consequences that partial penetration of light has for the spontaneous deformation of thin flat bodies \cite{corbett:deep,korner:nonlinear}.

Actually, a non-monotonic effect of light intensity upon deformation was also found in \cite{corbett:deep}, but it was predicted to vanish gradually for very large intensities, not with the abrupt decay shown here. We have found in our simplified approach that very same lack of monotonicity, which can have important consequences in applications. This establishes that partial penetration of light is not solely responsible for it.

\begin{comment}
\subsection{Lack of monotonicity in the intensity-deflection relation.}

\subsection{Optimal intensity, for given ribbon's length.}

\subsection{Critical value of the length (depending only on mu), below which no deflection whatsoever takes place.}

\subsection{Limitation of the model: it does not account for partial penetration of light.}

\end{comment}

\backmatter

%\bmhead{Acknowledgments}
%Acknowledgments are not compulsory. Where included they should be brief. Grant or contribution numbers may be acknowledged.

\begin{appendices}

\section{About the Corbett-Warner Model}\label{sec:appendix}
In this Appendix, which  has a pedagogical character, we give details about the statistical model by Corbett and Warner presented in Sect.~\ref{sec:model}. Its contents are derived, with minor modifications and adaptations, from \cite{corbett:polarization} and \cite{bai:photomechanical} (see also \cite{singh:model}).
\subsection{Step tensors}\label{sec:step_tensors}
A polymer strand in the reference configuration is represented as a chain of $N$ rigid rods, each of length $a$, freely jointed one to the adjacent ones, so that the orientation $\uv_i\in\sphere$ of the $i$th rod is completely independent from the orientation $\uv_j\in\sphere$ of any other rod with $j\neq i$.\footnote{It should perhaps be recalled that both photoresponsive molecules in the \ts\ configuration and photoinert mesogens are treated on the same footing in the reference configuration.} 

The \emph{span vector} vector $\Rr$ joining the ends of a strand is thus defined as
\begin{equation}
	\label{eq:span_vector_reference}
	\Rr:=a\sum_{i=1}^N\uv_i
\end{equation}
and the \emph{step tensor} $\Lr$ is correspondingly given by
\begin{equation}\label{eq:step_tensor_defintion}
	\Lr:=\frac{3}{Na}\ave{\Rr\otimes\Rr},
\end{equation}	
where the brackets $\ave{\cdots}$ denote ensemble averaging, as in the main text. By the mutual independence of rods, also in view of \eqref{eq:Q_definition} written for $\Qr$ and \eqref{eq:step_tensor_defintion}, we readily arrive at 
\begin{equation}
	\label{eq:step_tensor_reference}
	\Lr=\frac{3Na^2}{Na}\ave{\uv\otimes\uv}=a(3\Qr+\I).
\end{equation}
Use of the uniaxial representation for $\Qr$ in \eqref{eq:step_tensor_reference} leads us straight to \eqref{eq:step_tensor_reference_text} and \eqref{eq:principal_chain_steps_reference}.	It is perhaps worth noting that for $\Qr=\bm{0}$, which represents the isotropic distribution, $\Lr=a\I$, which accordingly represents a globule of radius $a$, thus justifying the scaling in definition \eqref{eq:step_tensor_defintion}.

To give the principal chain steps $(\lcper,\lcpar)$ featuring in \eqref{eq:step_tensor_current_text} the expressions in \eqref{eq:principal_chain_steps} we must recall that in the present configuration rods of different lengths coexist in one and the same polymer strand, while obeying different statistics: photoinert rods of length $a$ are in the nematic phase, whereas photoactivated rods of length $b<a$ are in the isotropic phase. Letting $\phi$ be the number fraction of the latter, there will be $(1-\phi)N$ rods of length $a$ and $\phi N$ rods of length $b$, so that the span vector $\Rc$ can be written as 
\begin{equation}
	\label{eq:span_vector_current}
	\Rc=a\sum_{i=1}^{(1-\phi)N}\uv_i+b\sum_{j=1}^{\phi N}\vv_j,
\end{equation}
where $\uv_i\in\sphere$ and $\vv_j\in\sphere$ are unit vectors along photoinert and photoactivated molecules, respectively. Now, $\uv_i$ and $\vv_j$ are clearly independent from one another, so that for
$\uv$ and $\vv$ representative of their ensembles,
\begin{subequations}\label{eq:averages}
\begin{equation}\label{eq:average_u_v}
\ave{\uv\otimes\vv}=\bm{0}.
\end{equation}
Moreover, the $\vv$'s are assumed to be distributed isotropically,
\begin{equation}
	\label{eq:average_v_v}
	\ave{\vv\otimes\vv}=\frac13\I,
\end{equation} 
and the $\uv$'s unixially,
\begin{equation}
	\label{eq:average_u_u}
	\ave{\uv\otimes\uv}=\Qc+\frac13\I,
\end{equation}
with $\Qc$ as in \eqref{eq:Q_unixial_representation}.
\end{subequations}
Making use of all equations \eqref{eq:averages} to evaluate the average $\ave{\Rc\otimes\Rc}$ from \eqref{eq:span_vector_current} and normalizing $\Lc$ to the length $Na$ of the unirradiated polymer, precisely as in \eqref{eq:step_tensor_defintion}, we readily arrive at equations \eqref{eq:step_tensor_current_text} and \eqref{eq:principal_chain_steps} in the main text.

\subsection{Equilibrium \cs-population}\label{sec:cis_population}
We follow \cite{corbett:nonlinear,corbett:polarization}, in the reinterpretation given in \cite{bai:photomechanical}, to calculate the number fraction $\phi$ of photoresponsive molecules in the \cs-state resulting from a dynamical equilibrium between forward and backward isomerizations.

Let $N$ be, as above, the total number of monomers in a polymer strand and let $A$ be the fraction of photoresponsive molecules among them. Thus, denoting by $\Nt$ the number of \ts-molecules and by $\Nc$ the number of \cs-molecules, conservation of mass requires that 
\begin{equation}
	\label{eq:mass_conservation}
	\Nt+\Nc=AN.
\end{equation}

The forward reaction rate $\ratetc$ is proportional to the product of $\Nt$ and the average projected light intensity along the molecular direction $\uv$,
\begin{equation}
	\label{eq:rate_t_c}
	\ratetc=\Gamma E^2\ave{(\uv\cdot\e)^2}_\mathrm{t}\Nt,
\end{equation}
where $\Gamma$ is a constant, $\bm{E}=E\e$ is the wave electric field, and the average $\ave{\cdots}_\mathrm{t}$ should only be computed on the \ts-molecules. However, since the latter obey the same statistics as all nematogenic molecules, the partial \ts-average is just the same as the full average,
\begin{equation}
	\label{eq:average_equality}
	\ave{(\n\cdot\e)^2}_\mathrm{t}=\ave{(\n\cdot\e)^2}.
\end{equation}

The thermally induced backward reaction has a rate simply proportional to $\Nc$,
\begin{equation}
	\label{eq:rate_c_t}
	\ratect=\frac{1}{\tau}\Nc,
\end{equation}
where $\tau$ is a thermal relaxation time. Equilibrium requires that $\ratetc=\ratect$. Inserting \eqref{eq:rate_c_t}, \eqref{eq:average_equality}, and \eqref{eq:rate_t_c} into this equality, we easily see that
\begin{equation}
	\label{eq:phi_formula}
	\phi=\frac{\Nc}{N}=A\frac{I\ave{(\e\cdot\n)^2}}{1+I\ave{(\e\cdot\n)^2}},
\end{equation}
where $I$ is as in \eqref{eq:relative_intensity} with
\begin{equation}
	\label{eq:intensities}
	\mathcal{I}=E^2\quad\text{and}\quad\mathcal{I}_\mathrm{c}=\frac{1}{\tau\Gamma}.
\end{equation}
To obtain \eqref{eq:phi_equation} from \eqref{eq:phi_formula}, it now suffices to observe that 
\begin{equation}
	\label{eq:observation}
	\ave{(\e\cdot\n)^2}=\e\cdot\ave{\uv\otimes\uv}\e=S(\n\cdot\e)^2+\frac13(1-S),
\end{equation}
where use has also been made of \eqref{eq:Q_definition} and \eqref{eq:Q_unixial_representation}.




\end{appendices}


%\bibliography{elastomers.bib}% common bib file




\end{document}